\documentclass[a4paper,oneside,english]{scrartcl}

\usepackage{amssymb}
\usepackage{amsmath}

\usepackage{graphicx}
\usepackage{epsfig}
\usepackage[utf8]{inputenc}

\usepackage[english]{babel}
\usepackage{cite}

\usepackage{geometry}
\begin{document}



\title{
Reduction of the electrodynamics of superconductors to those for conductors with the incorporation of spatial dispersion
}

\author{M.~A. Dresvyannikov\footnote{P.N. Lebedev Physical Institute of RAS, Leninsky pr. 53, 119991 Moscow, Russia},
				N.~A. Volchkov\footnotemark[1],
				A.~L. Karuzskii\footnotemark[1]   \footnote{E-mail:karuzskijal@lebedev.ru}, \\
				V.~A. Kulbachinskii\footnote{M.V. Lomonosov Moscow State University, 119991 GSP-1, Moscow, Russia},
				Yu.~A.Mityagin\footnotemark[1],
				A.~V. Perestoronin\footnotemark[1], \\
				A.~P. Chernyaev\footnote{Moscow Institute of Physics and Technology (State University), Institutsky Per. 9, Dolgoprudny, Moscow District 141700, Russia} } 








\date{}

\maketitle

\begin{abstract}
The study derives general frequency dependencies of the surface impedance modulus for conductors without the dc dissipation, i. e. for superconductors or 
perfect conductors. The frequency-dependent surface impedance was applied for the solutions corresponding to the spatially dispersive eigenvalues of the permittivity operator for conductors.
General reasons of formal dissipation-dependent effects of nonlocality in the electrodynamics of the surface impedance for conductors are 
considered with an emphasis on the analysis of nonself-adjoint problems.
The study demonstrates
that appropriately taken into account effects of the spatial dispersion can give the general frequency dependence of the surface impedance 
for the obtained solutions including that
for a superconductor.
It is shown that
incorporation
of the spatial dispersion leads to an appearance  of the Meissner effect in perfect conductors in the same manner as in superconductors.
%
%
%
The conclusion 
is obtained technically  for the first time from
the zero-frequency limit of the penetration depth
by substituting the surface impedance 
in the expression of ponderomotive Abraham force.
%
%
%
%
%
\end{abstract}

\section{Introduction}

Recently the frequency-dependent surface impedance was calculated for the spatially dispersive eigenvalues of the permittivity operator in conductors \cite{Dresvyannikov}.
It is reasonable to give
a brief description of
these formulations valid for both conductors and superconductors.
The electrodynamics of superconductors 
is supposed
to be in principle reduced to those for conductors
as the temperature approaches the critical temperature $T_{c}$  and beyond $T_{c}$.
However, such  a reduction
is not rather straightaway. Early the problem resulted in a supplement of the Maxwell equations by postulating 
additional, London equations for an explanation the Meissner effect \cite{SuperfluidsI,mende}.
It discriminates the perfect conductor in the non-zero-field cooling
behaviour
as compared to the superconductor. In a perfect conductor the Maxwell 
equations allow the static, spatially homogeneous field solutions \cite{SuperfluidsI,mende} below $T_{c}$. 
The
allowed spatially homogeneous penetration of the field
in the
cooled perfect conductor 
contradicts with the empirically observed 
static field expelled out of a superconductor.
To avoid this contradiction the Meissner effect was introduced in models
and theories
of the superconductivity  as the second necessary
property of the superconducting state in addition to the first
necessary requirement of the perfect conductivity \cite{SuperfluidsI,mende}. The postulation of additional, London equations is an example of
this introduction.




It is well known that the non superconducting skin effect can be normal or anomalous \cite{lifshiz,SilRuh,Abrikos,mende,Aude}. In the case of anomalous
skin effect, one have to solve the kinetic equation with the boundary conditions and with the incorporation of the spatial dispersion to find the response
of Fermi liquid on electromagnetic field \cite{lifshiz,SilRuh,Abrikos,mende,Aude,lifshiz2}. Necessity to involve the spatial dispersion in the kinetic equation follows from
the fact that one cannot neglect a nonlocal response of an electrical current on an electric field, since the
penetration depth of the electric field becomes smaller than the mean free path of electrons \cite{lifshiz,SilRuh,Abrikos,mende,Aude}. 
Probably the most illustrative
example of such 
an immediate nonlocal response of a particle in 
a certain spatial point 
on the electric field in 
an another spatially distant point is 
a movement of the solid. 
When the solid is considered as a rigid medium and moves as a whole, the 
force acting in the mentioned above spatially distant point 
produces a movement of the total body due to its rigidity, i. e. a movement of 
each spatial point of the body including the 
selected above
certain spatial point. It means an appearance in the spatially dispersive 
electronic liquid 
the property resembling the rigidity of solids, but 
in the one spatial dimension determined by the geometry of the surface impedance approximation. 
Here this rigidity is considered in a general formal manner to compare the 
electrodynamics behaviour 
of the superconductor 
and the normal conductor. 

It may look simple but at the same time important that the nonlocality of the problem is
inseparable from its inhomogeneity which implies the presence of boundary of the metal. Nevertheless the inseparability is not rigorous. The response
of Fermi liquid on electromagnetic field in the case of normal
skin effect is local but inhomogeneous in the vicinity of the boundary  \cite{lifshiz,SilRuh,Abrikos,mende,Aude,lifshiz2}. On the contrary the perfect conductor in non-zero-field cooling behaviour allows the solution of Maxwell equations with spatially homogenous fields in spite of the presence of boundary of the metal  that contradicts with the Meissner effect \cite{SuperfluidsI,mende}. So the correct relation between the nonlocality of the problem and its inhomogeneity 
must consider dissipation-dependent conditions additionally to 
that which implies the presence of boundary of the metal.

An anomalous skin effect is usually considered in details in many textbooks and monographs 
which review the theoretical
and experimental 
studies of the surface impedance of superconductors (see for example \cite{lifshiz,SilRuh,Abrikos,mende,Aude}), because of its inherent relation to the problem of the surface impedance of superconductors.
Both the anomalous skin effect in an extremely anomalous limit and the skin effect in superconductors correspond to current carriers with the infinite length of the mean free path between collisions and differ in other characteristic 
parameters. Clear speculations explaining this difference and its formal general origins are presented here.

For an explanation of the microwave spatial effects a  microscopic kinetics theory  is used to obtain the constitutive equation that  relates a current density and a field in a metal,
and is 
developed on a rigorous base of the linearized Boltzmann equation \cite{lifshiz,SilRuh,mende,Abrikos,Aude,Reuter}.
One can find lots of formulae which use the spatial nonlocality of the dielectric permittivity. 
However, 
according to the comments 
of \cite{mende}
these expressions are complicated in frames of the rigorous microscopic model due to the sophistication of a theoretical description and do not provide an illustrative and clear physical 
view of processes taking place in a metal.
Obviously it enhances a peril of losses of additional or novel wave solutions of the microwave kinetics problem.

Really the kinetics, i. e. non-equilibrium approach leads to the electrodynamic problem with 
a dissipation
that implies the problem with a nonself-adjoint \cite{ilyin1983,Shilov} differential operator.
A theory of nonself-adjoint operators is in the stage of a development now and is far from that degree of a completion, which corresponds to a theory of self-adjoint operators \cite{ilyin1983}. Since the wave problems of the anomalous skin effect in a metal and the skin effect in a superconductor belong to the strongly nonself-adjoint electrodynamic problems, but not to the 
weak disturbance of self-adjoint operators \cite{ilyin1983}, the mentioned peril of losses of solutions is actual. Nonself-adjoint problems may possess unusual properties, for instance, may have eigenfunctions without properties of the base functions \cite{ilyin1983}.
To exclude 
the peril of losses of the solutions additional to the already known solutions \cite{Reuter} 
it is important to analyse the possibility of an existence of additional solutions for this kinetics problem or of the alternative general approach to resolve the problem.
As an example, the original quantitative kinetics theory 
of the anomalous skin effect in a metal \cite{Reuter} 
does not
implicate the skin effect in a superconductor.
An appearance of additional solutions (additional waves) can strongly influence the 
properties of novel superconducting materials or nanoelectronic devices in the form of anomalous losses \cite{Shcherbakov2}, for example.

Microwave effects of the spatial dispersion have been considered recently in
 \cite{KarLT26,VolchKarPer} and 
rather general relation has been established 
between the eigenvalue of the
absolute permittivity operator and the surface impedance. 
It has been found 
in the surface impedance
\begin{math}
   (\widetilde{Z})
\end{math}
approximation  that the complex number
\begin{math}
        \widetilde{\varepsilon}_{aZ}
        =
        \mu_{0}
        \widetilde{Z}^{-2}
\end{math}
is the 
eigenvalue
of the absolute permittivity operator
\begin{math}
          \widehat{\widetilde{\varepsilon}}_{a},
\end{math}
where
\begin{math}
        \mu_{0}
\end{math}
is the vacuum permeability. 
Here and below the sign tilde denotes a complex value. The 
found relation 
corresponds with the usual expression for a transverse wave propagating into a conductor 
\begin{math}
   \widetilde{Z}=(\mu_{0}/\widetilde{\varepsilon}_{a})^{1/2}
\end{math}
in accordance with the plane wave geometry of the surface impedance
approximation \cite{mende,lifshiz2,KarLT26}.

This finding leads to the statement of a general wave problem, which is formulated to search the possible eigenvalues
\begin{math}
  \widetilde{\varepsilon}_{a}
\end{math}
of the absolute 
permittivity operator
\begin{math}
          \widehat{\widetilde{\varepsilon}}_{a}.
\end{math}
The statement is based on the known conclusion that an operator of the dielectric permittivity alone 
defines completely a microwave response of conductors with the spatial dispersion \cite{lifshiz2,lifshiz,SilRuh}. 
The eigenvalue problem 
has been formulated \cite{KarLT26} 
similar to the problem of the wave propagation in hollow waveguides and resonators, 
but for the nonself-adjoint permittivity operator. 
It relates with 
the impedance of a transverse wave
propagating out of a boundary surface into a conductor.
The corresponding solutions are the stable waves for the constitutive equation in a self-consistent microwave field due to the problem formulation.
The operator
\begin{math}
          \widehat{\widetilde{\varepsilon}}_{a }
\end{math}
is determined by an infinite system of equations of motion for individual particles together with the field equations \cite{lifshiz2,lifshiz,SilRuh}.
For the description of such systems the probabilistic methods are used and the kinetic equation for the distribution function with the self-consistent field should be solved.
Moreover kinetics problems with the spatial dispersion are nonlinear on spatial parameters because of the self consistent field coming into the constitutive equation  \cite{lifshiz2,lifshiz,SilRuh,KarLT26}.

The spatial dispersion, characterized by
the nonlocal response of an electrical current
on an electric field, 
results in the dependence 
of the components of 
permittivity
on the spatial parameter \cite{lifshiz2,lifshiz,SilRuh}.
Appropriate
spatial components of the  eigenvalue of permittivity
should be included in the
\begin{math}
   \widetilde{Z}=(\mu_{0}/\widetilde{\varepsilon}_{a})^{1/2}
\end{math}
for a given wavenumber
\begin{math}
    k 
\end{math}
due to the surface impedance approximation.
It has been shown that
the permittivity 
is a function
of the free parameter
\begin{math}
    k
\end{math}
in the homogeneous electronic liquid
\cite{KarLT26}.
This free parameter eventually depends on characteristics of an incident field and on dissipation-dependent effects of nonlocality,
as 
considered in detail here.

To select the 
spatial mode of the permittivity spectrum
one needs to
take 
into account \cite{KarLT26} 
that properties of the medium in the case of the prevalence of the spatial dispersion over the frequency dispersion
correspond to the retardation of the phase velocity of electromagnetic waves
\begin{math}
   v_{ph}= \omega/k \rightarrow 0,
\end{math}
according to  \cite{lifshiz2,lifshiz,SilRuh}.
The inhomogeneous in space field can be considered as static in comparisons with the reaction of the conducting medium, characterized by the Fermi velocity $v_{F}\sim0.01\,c$, where $c$ is the speed of light. It has been argued 
that the 
response to the external influence becomes spatially non-local, and the
state of the system is determined by the correlation of the applied forces \cite{KarLT26}. In the static and quasistatic consideration, the maximum perturbation of the conducting medium should
be caused by that of possible spatial configurations of the medium, which would correspond to the influence of the dominating acting ponderomotive force 
produced by the incident field. 

The ponderomotive force (per unit volume) can be calculated \cite{KarLT26} by the Maxwellian field-stress tensor, each component of which is the spatial density of the corresponding momentum flux
\begin{math}
   \widetilde{\mathbf{\Pi}} = [ \widetilde{\mathbf{D}} \times \widetilde{\mathbf{B}}  ],
\end{math}
where
\begin{math}
   \widetilde{\mathbf{D}}
\end{math}
and
\begin{math}
    \widetilde{\mathbf{B}}
\end{math}
are the electric and  magnetic inductions.
Derivative
\begin{math}
    \partial \widetilde{\mathbf{\Pi}}/\partial t
\end{math}
determines the ponderomotive Abraham force
\cite{lifshiz2,SilRuh}.
The spatial-type force resonance corresponds to a pure real value or to a pure imaginary value of the expression that determines the ponderomotive Abraham force, with the canceled temporal factor.
Previously it has been demonstrated \cite{Dresvyannikov,KarLT26} that
a substitution of the eigenvalue of the permittivity operator into  the  Maxwellian field-stress tensor \cite{lifshiz2,SilRuh}
brings to conditions of spatial-type force resonances.
Substituting the impedance value  into the momentum flux  equation one can deduce \cite{Dresvyannikov,KarLT26}
that
the spatial-type force resonances correspond to a pure real or to a pure imaginary value of the coefficient
\begin{math}
   (
   \mu_{0}^{1/2} \widetilde{\varepsilon}^{\,3/2}
   )
\end{math}
in the obtained expression.
Then from this relation the conditions for spatial-type force resonances 
have been found in the form of certain values of 
an argument of 
the complex surface impedance.  
These deduced formally spatial-type force resonances correspond to 
the dominating 
ponderomotive force with the respective spatial configuration. 
Those include a particular solution corresponding to a traditional 
anomalous skin effect in a conductor,
to a surface impedance of a superconductor and 
several additional resonance solutions, which have not yet been analysed in detail \cite{Dresvyannikov,KarLT26}.


Here the study demonstrates
that appropriately taken into account effects of the spatial dispersion can give  general frequency dependencies of the surface impedance 
for the obtained in \cite{Dresvyannikov,KarLT26} solutions including those both for the normal conductor and for the superconductor or perfect conductor.
General reasons of formal dissipation-dependent effects of nonlocality in the electrodynamics of the surface impedance for conductors are 
considered with an emphasis on the sufficiently abstract analysis of nonself-adjoint problems presented 
mainly in the Sec. \ref{Formal backgrounds}.

It is shown that
incorporation
of the spatial dispersion leads to an appearance  of the Meissner effect in perfect conductors in the same manner as in superconductors.
%
%
The conclusion 
results technically from an evaluation of the indeterminate form in the zero-frequency limit following  \cite{lifshiz2,SilRuh}, however,
for the first time 
by substituting the surface impedance 
in the expression of Abraham force. Fortunately such a substitution leads to the correct values of the magnetic field penetration depth as follows from comparisons with the substitution of 
London equations in the Sec. \ref{London equations and Abraham force}. 
%
%
More detailed considerations of the evaluation procedure for this indeterminate form 
demonstrate that the process of
finding the proper, empirically verified solution of the electrodynamic problem may require an introduction of the effective constitutive parameters 
and pairing, or a postulation of the additional phenomenological equations 
depending on the kind of representation of the original problem operator. 
%
%
This conclusion is supported by different values of the permeability constant derived from an evaluation of 
the zero-frequency limit for 
a prevalence of the spatial dispersion or of the frequency dispersion
\cite{lifshiz2,SilRuh}. There it has been found that the empirically verified value corresponds to the spatial dispersion.
It is noted here that formal 
effects of the electrodynamic nonlocality 
for the nonself-adjoint problems can affect the linear properties and commutation of acting operators. 
%
%
The found 
dependence on the operator representation can be the result of the lost linearity and commutation, especially in the case of evaluating the indeterminate form in the zero frequency limit. 


The further usefulness of the goal stated here
is in the approach that 
has been used to reach the goal. The 
deduced formally spatial-type force resonances provide a formalism that decreases significantly a peril of loses of additional solutions of the electrodynamics problem.

%
The text
presents an analysis of
a particular physical system,
a
well-known
perfect conductor \cite{SuperfluidsI,mende}, for which it
is possible to show the advantages of
the developed
formal method.
The criterion would have to be just a demonstration of
its adequacy of any physical system with a comparison of the application
of the ordinary Maxwell's equations and the kinetic equation.

%
%

In the problem considered 
it elaborates the possibility of Meissner effect in perfect conductors. The method demonstrates that incorporation
of the spatial dispersion leads to an appearance of the Meissner effect in perfect conductors in the same manner as in superconductors.
The 
surface impedance approximation
comprehensibly shows here that the perfect conductivity  results
in inevitable expelling the field, the Meissner effect, when
the correlation of
ponderomotive Abraham forces is taken into account as
an effect of the spatial dispersion.
It allows to
refuse the
mentioned above introduction of the Meissner effect as the second necessary
attribute of the superconducting state in models
of the superconductivity 
when the first
necessary requirement
is the perfect conductivity \cite{SuperfluidsI,mende}.
The conclusion
is derived without
the
application of the kinetic equation. It means that the conclusion does not require details of the microscopic structure of the system and has the most general
character in frames of the used approximation.

Moreover the attribution of the Meissner effect to the perfect conductivity, but not only
to
the superconductivity,
expands the
diversity of materials, systems or 
devices, which can possess the superconducting-like properties.
Possible examples of such systems are briefly referred at the end of the Sec. 
\ref{Effective constitutive parameters}.







\section{Frequency dependencies} 
\label{Frequency dependence}

To evaluate the general frequency dependence of the eigenvalue modulus
\begin{math}
    | \widetilde{\varepsilon}_{a} |
\end{math}
of absolute permittivity operator
\begin{math}
    \widehat{\widetilde{\mathbf{\varepsilon}}}_{a}
\end{math}
and of the surface impedance modulus
\begin{math}
    | \widetilde{Z} |,
\end{math}
let us again  \cite{Dresvyannikov} consider at first only the spatial effects  in permittivity, assuming the problem is stationary 
\begin{math}
    \omega = 0
\end{math}
and taking into account only the spatial field 
inhomogeneity in the form of a wave number
\begin{math}
   k''.
\end{math}
Preliminary results concerning the frequency dependencies and spatial effects 
were discussed in \cite{Dresvyannikov-2016-SPIE,Dresvyannikov-2018-arXiv,VolchKarPer-2018-JPCS,Dresvyannikov-2018-PSIJ}.

\subsection{Normal conductivity}

The first of Maxwell equations 
can be written  as
\begin{equation}
          \label{MaxwellEq:curr}
           \mathrm{rot}\; \widetilde{\mathbf{H}}
          =
          (\partial \widetilde{\mathbf{D}}/ \partial t)
          +
          \widetilde{\mathbf{j}}.
\end{equation}
Below the vectorial values of electromagnetic field are meant as the eigenfunctions 
compliant with the eigenvalue of permittivity operator
\begin{math}
    \widehat{\widetilde{\varepsilon}}_{a}
\end{math},
i. e. with the numerical complex value
\begin{math}
          \widetilde{\varepsilon}_{a}
\end{math}
in the 
constitutive equation \cite{Dresvyannikov} 
\begin{math}
  \widetilde{\mathbf{D}}
          =
          \widehat{\widetilde{\varepsilon}}_{a} \widetilde{\mathbf{E}} 
          =
          \widetilde{\varepsilon}_{a} \widetilde{\mathbf{E}} .
\end{math}
The frequency
\begin{math}
    \omega''_{k}
    =
    k'' v_{F}
\end{math}
can be associated with the spatial 
inhomogeneity
\begin{math}
    k''
\end{math}
 \cite{Dresvyannikov}, where
\begin{math}
     v_{F}
\end{math}
is the Fermi velocity, i. e. the velocity of  propagation of the external perturbation.
Using the ambiguity of the representation of a right side of the previous expression \cite{lifshiz2,lifshiz,SilRuh} 
and integrating it over the time one can get  from the constitutive equation \cite{Dresvyannikov} 
\begin{math}
           \widetilde{\mathbf{D}}\left( \mathbf{r}, t \right)
          =
           \widetilde{\varepsilon}_{a}
          \widetilde{\mathbf{E}}\left( \mathbf{r}, t \right)
\end{math}
an expression \cite{KarLT26}
\begin{equation}
          \label{PermittivityConductivity}
          \widetilde{\varepsilon}_{a}
          =
          \widetilde{\varepsilon}_{p}
          \varepsilon_{0}
          -
          i \widetilde{\sigma} / \omega''_{k}
          =
          \widetilde{\varepsilon}_{p}
          \varepsilon_{0}
          -
          i \widetilde{\sigma} / ( k'' v_{F}),
\end{equation}
where
\begin{math}
          \widetilde{\sigma}
\end{math}
is the static eigenvalue of
conductivity operator,
\begin{math}
     \widetilde{\varepsilon}_{p}
\end{math}
is the eigenvalue of relative permittivity operator of a lattice. 
%
%
This formula, together with the complex 
argument conditions of the spatial
resonances and the plain
geometry of the surface impedance, uses the property of spatial nonlocality of the dielectric permittivity.
%
%
The solutions found in  \cite{Dresvyannikov,KarLT26} for conductors 
and superconductors have 
the 
fixed arguments of complex numbers
\begin{math}
          \widetilde{\varepsilon}_{a}
\end{math}
and 
\begin{math}
          \widetilde{Z}\sim1/\widetilde{k}
\end{math}
determined by 
conditions of the spatial force resonances. 
So for a finite frequency
\begin{math}
          \omega
\end{math}
of an external field the spatial 
inhomogeneity
\begin{math}
          k''
\end{math}
is determined, according to \cite{Dresvyannikov,KarLT26}, 
by the frequency and by the eigenvalue of 
permittivity operator
\begin{equation}
        \label{DispersionLaw}
          k''
          \sim
          \omega ( \widetilde{\varepsilon}_{a}  \mu_{0}  )^{1/2}
          =
          \omega /\widetilde{v}_{ph},
\end{equation}
that may be assigned to a mean field approximation. The conductivity in the extremely anomalous limit is determined by the frequency
\begin{math}
          \widetilde{\sigma}
          \sim
          1 / \omega
\end{math}
also \cite{VolchKarPer}.
The substitution of
\begin{math}
          k''
\end{math}
and
\begin{math}
          \widetilde{\sigma}
\end{math}
into the expression (\ref{PermittivityConductivity}) for
\begin{math}
          \widetilde{\varepsilon}_{a}
\end{math}
results in 
the proportionality
\begin{math}
          ( \widetilde{\varepsilon}_{a}    )^{3/2}
          \sim
          \omega^{-2}
\end{math}
that corresponds to the frequency dependence of the impedance modulus
\begin{math}
          | \widetilde{Z} |
          \sim
          \omega^{2/3}
\end{math}
for all solutions found for the conductor except that for the superconductor, as it was shown in  \cite{Dresvyannikov}. 

\subsection{Superconductors and perfect conductors}
\label{Superconductors and perfect conductors}
To estimate the frequency dependencies of moduli
\begin{math}
    | \widetilde{Z} |
\end{math}
and
\begin{math}
    | \widetilde{\varepsilon}_{a} |
\end{math}
for the solution found for superconductors or perfect conductors having the phase of
\begin{math}
          \widetilde{\varepsilon}_{a}
\end{math}
equal to
\begin{math}
          \beta_{1}=\pi
\end{math}
and the phase of
\begin{math}
          \widetilde{Z} 
\end{math}
equal to
\begin{math}
          \psi_{1}=\pi/2
\end{math}
(and additionally for the solution with
\begin{math}
          \alpha_{0}=0
\end{math}
and
\begin{math}
          \varphi_{0}=0)
\end{math}
\cite{Dresvyannikov} one should note that for these solutions the real parts of  eigenvalue
\begin{math}
          \widetilde{\sigma}
\end{math}
of the conductivity operator in the equation (\ref{PermittivityConductivity})
equal zero and the direct current does not dissipate with time.

Spatial dispersion effects originate from the action produced by an electric field during the "free" motion of a particle along the not too long segments of its trajectory 
\cite{lifshiz}. The order of the magnitude of this length of the mean free path is determined by two mechanisms: collisional 
scattering, which limits the free motion of a carrier along its trajectory, and averaging 
an oscillating field during the mean free time spent by a carrier in its movement along the trajectory between successive collisions \cite{lifshiz}. The 
least 
of these two 
lengths of segments of the trajectory confines the upper limit of the length of the mean free path 
and 
dominates in the spatial dispersion effects.

Without the dc dissipation the collisional scattering is absent and the spatial dispersion is determined by the spatial distribution of the field.
In this case, one should take into account an influence of spatial
inhomogeneities
of the field with the 
smallest length, regarding an absence of the dc dissipation, for all possible spatial 
scales, but not only for those corresponding to the frequency of an incident wave 
(cf. the Eq. (\ref{DispersionLaw})) in the presence of the nonzero dc dissipation. 
The smallest length corresponds to the largest possible value of a wave number
\begin{math}
    k''
\end{math}
or of an associated frequency
\begin{math}
    \omega''_{k}
    =
    k'' v_{F}.
\end{math}
The velocity of propagation of a perturbation
in the
quasi-stationary
environment without 
the dc dissipation could correspond to
the 
spatially dispersive value
\begin{math}
     v,
\end{math}
which should not be necessarily associated with the Fermi velocity
\begin{math}
     v_{F},
\end{math}
but 
could be equal to the speed of light
\begin{math}
     c,
\end{math}
for example, as in  the  Langmuir plasma frequency. 
The 
highest value
of spatial
inhomogeneities
of the field with the largest possible value of a wave number
\begin{math}
    k''
\end{math}
or of an associated frequency
\begin{math}
    \omega''_{k}
    =
		k'' v,
\end{math}
in this case,
can approach the vertical-asymptote values 
equal to the
\begin{math}
    k''_{0}
\end{math}
or 
to the frequency proportional to
\begin{math}
    \omega''_{kp0}
    =
    k''_{0} v_{F},
\end{math}
 corresponding to the  Langmuir plasma frequency
\begin{equation}
        \label{LangmuirPlasmaFrequency}
    \omega_{p}
    =
    k''_{0} c
    =
    c/\lambda_{0},
\end{equation}
where
\begin{equation}
        \label{LondonPenetrationDepth}
    \lambda_{0}
    =
    (k_{0}'')^{-1}
    =
    (m/\mu_{0}ne^{2})^{1/2}
\end{equation}
is the London 
penetration depth at the zero temperature (the plasma penetration depth).
The absence of the dc dissipation should result in the feasibility of "acoustic" conditions 
\begin{math}
    (  v >0)
\end{math}
for the spatially dispersive plasmon 
polariton frequency
\begin{math}
    \omega''_{kp}
    =
    k'' v. 
\end{math}

This spatial 
inhomogeneity will correspond to the highest value of the Abraham force  (see the Eq.(\ref{AbrahamForce2}) below and Ref. \cite{Dresvyannikov}). So it should result in the spatial structure dominating over the others possible structures \cite{Dresvyannikov,KarLT26}. 
A value of this spatial 
inhomogeneity will not depend on frequency values
\begin{math}
    \omega
\end{math}
of any incident microwaves, while those could be considered as
quasi-stationary
when
\begin{math}
    \omega
    <
    \omega''_{kp0}
    \ll
    \omega_{p}.
\end{math}
This can be illustrated by an expression of a current density after the end of an external perturbation obtained with the account of the spatial dispersion \cite{VolchKarPer}.
If one substitutes the spatial factor in Eq.(3) of Ref.\cite{VolchKarPer} in 
the form of
\begin{equation}
        \label{StaticCurrentDensity}
    j_{0}(\theta,r)
    =
    j_{0}(\theta)e^{i\kappa r}, 
\end{equation}
where
\begin{math}
    \theta
\end{math}
is the reduced temperature
\begin{math}
    \theta
    =
    T
    /
    T_{c}
\end{math}
\cite{VolchKarPer,mende}, it is reexpressed as the current density
\begin{equation}
        \label{CurrentDensity}
    j(\theta,r,t)
    =
    j_{0}(\theta)e^{ -\omega ^{\prime}_{k}t }
    e^{i(\kappa r \pm \omega^{\prime\prime}_{k}t)}, 
\end{equation}
which remains after switching off an electric field
\begin{math}
    E=E_{0}e^{i(kr+\omega t)}.
\end{math}
In a perfect conductor the direct current does not decay with time
\begin{math}
    (\omega ^{\prime}_{k}
    =
		0),
\end{math}
and the stationary current solution (\ref{StaticCurrentDensity}) may be reconstructed from (\ref{CurrentDensity}) by the proper time averaging of plasma oscillations with positive and negative signs of the 
plasmon-polariton
frequency 
\begin{math}
    \omega''_{k}
    =
    \omega''_{kp0},
\end{math}
by a finite time shift in their starting moments, for example.

The requirement of an independence of the spatial 
inhomogeneity
\begin{math}
    k''
\end{math}
on the frequency
\begin{math}
    \omega
\end{math}
should result, according to the equation 
\begin{equation}
          \label{DispersionLaw2}
          k''
          =
          \omega ( | \widetilde{\varepsilon}_{a} |  \mu_{0}  )^{1/2}
          =
          \omega / | \widetilde{v}_{ph} |,
\end{equation}
in the inverse ratio 
\begin{math}
    \sqrt{ | \widetilde{\varepsilon}_{a} | \vphantom{ X^{2} } }\sim 1/\omega
\end{math}
and in the dependence of the surface impedance (
its modulus) linear on a frequency
\begin{math}
    \omega
\end{math}
\begin{equation}
    \label{SurfaceImpedanceOnFrequency}
    | \widetilde{Z} |
    =
    \sqrt{\frac{\mu_{0}}{| \widetilde{\varepsilon}_{a}|}}
    =
    \frac{\omega\mu_{0}}{k''}.
\end{equation}
It may look surprising 
that 
similar equations (\ref{DispersionLaw}) and (\ref{DispersionLaw2}) produce different solutions. However, this is not an
error,
but those are two different possible formal solutions of this equation following from different conditions of spatial dispersion.  
The frequency dependence of the eigenvalue
\begin{math}
     \widetilde{\varepsilon}_{a}
\end{math}
in the right side of the Eq. (\ref{DispersionLaw2}) is determined by the requirement of 
 the 
 independent on
\begin{math}
    \omega 
\end{math}
wavenumber
\begin{math}
          k''
\end{math}
in the left side, 
inversely to their correlations 
in the Eq. (\ref{DispersionLaw}), 
where the wavenumber
\begin{math}
          k''
\end{math}
is determined by the eigenvalue
\begin{math}
     \widetilde{\varepsilon}_{a}
\end{math}
according to the mean-field approximation.

Presented here formal considerations of the dissipation-dependent spatial effects illustrate the mentioned above feasibility of "acoustic" conditions for intrinsic 
excitations 
in the absence of the dc dissipation. The plasma penetration depth (\ref{LondonPenetrationDepth}) does not implicate the dissipative parameters 
such as the mean free time, for example. It 
describes the synchronous, collisionless movement of current carriers with the concentration
\begin{math}
          n,
\end{math}
the charge
\begin{math}
          e,
\end{math}
and the mass
\begin{math}
          m 
\end{math}
that corresponds to the current density
\begin{math}
    j
    =
    env
\end{math}
for screening a stationary field \cite{SuperfluidsI,lifshiz2,SilRuh,VolchKarPer,mende}. 
The absence of the dc dissipation means that intrinsic elementary
excitations, which can lead to the dissipation, are not excited in the system of current carriers. The plasma penetration depth satisfies this requirement.
Instead of the 
dissipative excitations the spatial dispersion predicts the domination of 
the highest possible value
of spatial
inhomogeneities
of the field, in the absence of the dc dissipation, 
and the respective largest possible value of a wave number 
\begin{math}
    k''_{0} 
\end{math}
or of an associated frequency proportional to
\begin{math}
    \omega''_{kp0}
    =
    k''_{0} v_{F}. 
\end{math}
Combined with the requirement of 
independence of the highest possible spatial 
inhomogeneity
\begin{math}
    k''_{0}
\end{math}
on the frequency
\begin{math}
    \omega ,
\end{math}
while 
it can be considered as
quasi-stationary
when
\begin{math}
    \omega
    <
    \omega''_{kp0}
    \ll
    \omega_{p}, 
\end{math}
this frequency-independent inhomogeneity 
manifests the presence of the critical minimal velocity for intrinsic elementary excitations in the system of current carriers. 
The velocity of the perturbation with the quasi-stationary frequency
\begin{math}
    \omega 
\end{math}
is given by the Eq. (\ref{DispersionLaw2}), where
\begin{math}
    k''
    =
    k''_{0}.
\end{math}
The independence of the
\begin{math}
    k''_{0}
\end{math}
on the frequency means that the velocity of the perturbation with the quasi-stationary frequency is 
not enough to produce the intrinsic elementary excitations in the system of current carriers.
It corresponds to the movement of the system as a whole, without the dc dissipation, i. e. 
to the movement of a superfluid "rigid" medium with the possibility of the frictionless superflow between the adjacent layers, when the velocity of the movement is lower than the critical velocity for intrinsic excitations in accordance with Landau criterion for superfluidity.

The further temperature issue of how the electrodynamics of superconductors can be 
reduced to those for conductors as the temperature approaches and beyond the critical temperature may be clearly stated using, e. g., the
generalised
two-fluid principles
\cite{VolchKarPer,mende} or rigorous
modelling
\cite{mende,Aude}.



\section{The spatial dispersion and Meissner effect in superconductors and perfect conductors}
\label{Meissner effect}




The nonlocal response of any material changes in the vicinity of the boundary, if the distance
to the boundary becomes comparable with the spatial scale of the nonlocality. It is
considered here conventionally in the plain geometry of the surface impedance and in
more abstract formulations in the Sec. \ref{Formal backgrounds}.
%
%
%
%
In 
the surface impedance approximation 
\cite{lifshiz2,lifshiz,SilRuh,KarLT26,mende,Aude} let it will be
the coordinate $z$ with the basis vector $\mathbf{k}$, which are directed into the conductor together with the unitary normal $\mathbf{n}$ to the boundary surface $z=0$. A monochromatic transverse field at the normal incidence to the plane surface of a conductor will be considered.
An electric field strength vector is directed tangentially along the $x$-axis
(\begin{math}
   \widetilde{\mathbf{E}}(z)=\widetilde{E}_{x}(z)\textbf{i}
\end{math}),
and a magnetic induction vector is directed along the $y$-axis
(\begin{math}
   \widetilde{\mathbf{B}}(z)=\widetilde{B}_{y}(z)\textbf{j}
\end{math}).
Here \textbf{i} and \textbf{j} are the basis vectors. The fields decrease into the depth of a conductor as 
\begin{math}
   \widetilde{B}_{y}(z)
          =
          \widetilde{B}_{y}(0) e^{-z/\widetilde{\delta}}
\end{math} and
\begin{math}
          \widetilde{E}_{x}(z)
          =
          \widetilde{E}_{x}(0) e^{-z/\widetilde{\delta}},
\end{math}
 where 
\begin{math}
   \widetilde{\delta}
\end{math} is the complex penetration depth, which defines the surface impedance by the relation 
\begin{math}
          \widetilde{Z}
          =
          i \omega \mu_{0} \widetilde{\delta}
\end{math}
\cite{lifshiz2,lifshiz,SilRuh,KarLT26,mende,Aude}, 
and $\mu_{0}$ is the permeability of vacuum.
Since the surface impedance relates the tangential components of fields on the boundary surface 
\cite{KarLT26,mende}
\begin{equation}
    \label{SurfaceImpedanceRatios}
    \widetilde{Z}
    =
    \frac{\widetilde{E}_{x}(0)}{\widetilde{H}_{y}(0)}
    =
    \frac{\widetilde{B}_{y}(0)}{\widetilde{D}_{x}(0)}    ,
\end{equation}
the zero limit of the surface impedance
\begin{math}
    \widetilde{Z}|_{\omega\rightarrow0}
    =
    0
\end{math}
at the zero frequency, 
deduced from the Eq. (\ref{SurfaceImpedanceOnFrequency}), means that the stationary tangential component of an electric field (or of a magnetic induction) should be equal zero
\begin{math}
    \widetilde{E}_{x}(0)|
    _{\omega\rightarrow0}
    =
    0
\end{math}
(or
\begin{math}
    \widetilde{B}_{y}(0)|_{\omega\rightarrow0}
    =
    0),
\end{math}
but with a possible finite value of
\begin{math}
    \widetilde{H}_{y}(0)|_{\omega\rightarrow0}
    \neq
    0
\end{math}
(or
\begin{math}
    \widetilde{D}_{x}(0)|_{\omega\rightarrow0}
    \neq
    0).
\end{math}
An alternative combination of a finite value of
\begin{math}
    \widetilde{B}_{y}(0)|_{\omega\rightarrow0}
\end{math}
with an infinite
\begin{math}
    \widetilde{D}_{x}(0)|_{\omega\rightarrow0}
\end{math}
is possible in situations when the direct current flows without dissipations.
Formally the Eq. (\ref{SurfaceImpedanceRatios}) in this case,
corresponds to the indeterminate form such as
\begin{math}
    0\cdot\infty ,
\end{math}
an evaluation of which is closely related to the Meissner effect. The remainder of this 
study is devoted to the problem of 
evaluating this indeterminate form 
by 
substituting 
it in the expression of Abraham force.


Let us consider a behaviour of the Abraham force \cite{lifshiz2,SilRuh}
in a perfect conductor (or in a superconductor) 
in the limit of zero frequency
\begin{math}
    \omega\rightarrow0.
\end{math}
From the expression for the spatial density of the momentum flux \cite{Dresvyannikov,KarLT26}, which is the corresponding component of the Maxwellian field-stress tensor,
\begin{equation}
    \label{ComponentOfMaxwellianFieldStressTensor}
   \widetilde{ \mathrm{\Pi} } = [ \widetilde{\mathbf{D}} \times \widetilde{\mathbf{B}}  ]
   =
   (
   \mu_{0}^{1/2} \widetilde{\varepsilon}_{a}^{\,3/2}
   )
   \widetilde{E}_{x}^{2}(0) \mathbf{k} ,
\end{equation}
where
\begin{math}
        \mathbf{k}
\end{math}
is the basis vector of the \textit{z}-axis, the Abraham force 
can be 
derived \cite{lifshiz2,SilRuh} neglecting the frequency dispersion:
\begin{equation}
\label{AbrahamForce}
\begin{split}
    \left( 1-\frac{1}{\widetilde{\varepsilon}} \right)
    \frac{ \partial
    \widetilde{ \mathrm{\Pi} }
    }{\partial t}
    =
    \left( 1-\frac{1}{\widetilde{\varepsilon}} \right)
    \frac{ \partial
    }{\partial t}
    [ \widetilde{\mathbf{D}} \times \widetilde{\mathbf{B}}  ]
    =
    \\
    =
    \left( 1-\frac{1}{\widetilde{\varepsilon}} \right)
    \frac{ \partial
    }{\partial t}
    \left[
    \widetilde{D}_{x}(0) e^{i\omega t} \mathbf{i}
    \times
    \widetilde{B}_{y}(0) e^{i\omega t} \mathbf{j}
    \right]
    =
    \\
    =
     2 i \omega
    \left( 1-\frac{1}{\widetilde{\varepsilon}} \right)
    e^{i 2 \omega t}
    \left[
    \widetilde{D}_{x}(0) \mathbf{i}
    \times
    \widetilde{B}_{y}(0) \mathbf{j}
    \right],
\end{split}
\end{equation}
where
\begin{math}
        \widetilde{\varepsilon}
        =
        \widetilde{\varepsilon}_{a} /
        \varepsilon_{0}
\end{math}
is the eigenvalue of 
relative permittivity operator. The  surface impedance (\ref{SurfaceImpedanceRatios}) defines the relation between electric and magnetic inductions
\begin{math}
    \widetilde{D}_{x}(0)
\end{math}
and
\begin{math}
    \widetilde{B}_{y}(0)
\end{math}
and may be substituted in the expression (\ref{AbrahamForce}) for the Abraham force.
Then  using the Eq. (\ref{SurfaceImpedanceOnFrequency}) for  the modulus 
of surface impedance 
and taking account of its phase
\begin{math}
          \psi_{1}=\pi/2
\end{math}

\begin{equation}
\label{AbrahamForce2}
\begin{split}
    2 i \omega
    \left( 1-\frac{1}{\widetilde{\varepsilon}} \right)
    e^{i 2 \omega t}
    \left[
    \frac{ \widetilde{B}_{y}(0) }{ \widetilde{Z} }  \mathbf{i}
    \times
    \widetilde{B}_{y}(0) \mathbf{j}
    \right]
    = \\ =
    2 i \omega
    \left( 1-\frac{1}{\widetilde{\varepsilon}} \right)
    e^{i 2 \omega t}
    \left[
    \frac{ -i k''_{A} }{ \omega\mu_{0} }  \widetilde{B}_{y}(0) \mathbf{i}
    \times
    \widetilde{B}_{y}(0) \mathbf{j}
    \right]
    = \\ =
     2 
    \left( 1-\frac{1}{\widetilde{\varepsilon}} \right)
    e^{i 2 \omega t}
    \left[
    \frac{ k''_{A} }{ \mu_{0} }  \widetilde{B}_{y}(0) \mathbf{i}
    \times
    \widetilde{B}_{y}(0) \mathbf{j}
    \right]
    =\\ =
     2 
    \left( 1-\frac{1}{\widetilde{\varepsilon}} \right)
    e^{i 2 \omega t}
     \mu_{0}
    \left[
    \mathrm{rot}  \mathbf{\widetilde{H} }(0)
        \times
    \mathbf{\widetilde{H} }(0)
    \right].
\end{split}
\end{equation}
The expression (\ref{AbrahamForce2}) shows that 
Abraham force at all frequencies
\begin{math}
    \omega
    \ll
    \omega_{p},
\end{math}
as well as in the stationary limit
\begin{math}
    \omega\rightarrow0,
\end{math}
achieves the highest magnitude 
when the spatial 
inhomogeneity
\begin{math}
    k''_{A}
\end{math}
gets the largest, frequency-independent value (see Sec. 2.2) corresponding to
the London (or plasma) penetration depth
\begin{math}
    \lambda_{0}
    =
    (k_{0}'')^{-1}
    =
    (m/\mu_{0}ne^{2})^{1/2}.
\end{math}
Such the spatial configuration of the fields related to the reconstruction of an electronic system, which corresponds to the highest magnitude of the Abraham force, is dominant due to the force correlations and the consequential decrease of the free energy \cite{Dresvyannikov,KarLT26}. 
And respectively, the absolutely unstable spatial configuration, according to the Eq.  (\ref{AbrahamForce2}), corresponds to the geometry of the spatially homogeneous field and to the zero value of the spatial 
inhomogeneity
\begin{math}
     k''_{A}
    =
    0
\end{math}
when the Abraham force is zero with no advantage in  the free energy. So the conclusions obtained here in the surface impedance approximation show that the spatial dispersion results in an appearance of the Meissner effect in perfect conductors in the same manner as in superconductors contrary to the 
preceding considerations \cite{SuperfluidsI}, which do not incorporate the spatial dispersion effects. A driving force of its appearance is the Abraham force.

It should be noted that this conclusion may be derived more formally, without  considerations of any frequency dependencies. 
Abraham force 
can be 
reexpressed as
\begin{equation}
\label{AbrahamForceStationaryApproximation:a}
\begin{split}
    \left( 1-\frac{1}{\widetilde{\varepsilon}} \right)
    \frac{ \partial
    \widetilde{ \mathrm{\Pi} }
    }{\partial t}
    =
    \left( 1-\frac{1}{\widetilde{\varepsilon}} \right)
    \frac{
    \partial
    }{
    \partial t
    }
    \left[
    \widetilde{\mathbf{D}}
    \times
    \widetilde{\mathbf{B}}
    \right]
    =
    \\
    =
    \left( 1-\frac{1}{\widetilde{\varepsilon}} \right)
    \left[
    \frac{
    \partial \widetilde{\mathbf{D}}
    }{
    \partial t
    }
    \times
    \widetilde{\mathbf{B}}
    \right]
    +
    \left( 1-\frac{1}{\widetilde{\varepsilon}} \right)
    \left[
    \widetilde{\mathbf{D}}
    \times
    \frac{
    \partial \widetilde{\mathbf{B}}
    }{
    \partial t
    }
    \right]
\end{split}
\end{equation}
For monochromatic fields the Maxwell equations
in a homogeneous medium with the spatial dispersion may be written \cite{Dresvyannikov,lifshiz2,lifshiz,SilRuh,KarLT26} as
\begin{equation}
           \mathrm{rot} \widetilde{\mathbf{B}}
          =
           \mu_{0}\mathrm{rot} \widetilde{\mathbf{H}}
          =
           \mu_{0}
					\frac{
					\partial \widetilde{\mathbf{D}}
					}{
					\partial t
					};
           \phantom{---}
           \mathrm{rot} \widetilde{\mathbf{E}}
          =
          -
					\frac{
					\partial \widetilde{\mathbf{B}}
					}{
					\partial t
					};
          \label{MaxwellEq:a}
\end{equation}
\begin{equation}
            \mathrm{div} \widetilde{\mathbf{D}}
          =
           \mathrm{div} \widetilde{\mathbf{E}}
          =
          0 ;
           \phantom{-------}
           \mathrm{div} \widetilde{\mathbf{B}}
          =
          0 .
          \label{MaxwellEq:b}
\end{equation}
Due to an absence of magnetic charges and an equality
\begin{math}
    \partial \widetilde{\mathbf{B}}
    /
    \partial t
    =
    0
\end{math}
in the second of Eqs (\ref{MaxwellEq:a})
in 
the stationary approximation
\begin{math}
    \omega
    =
    0
\end{math}
the second term in the sum (\ref{AbrahamForceStationaryApproximation:a}) equals zero.
Regarding the ambiguity of the representation of the first Maxwell equation 
(cf.~the~first~of~Eqs~(\ref{MaxwellEq:a}) and the Eq.(\ref{MaxwellEq:curr})
\begin{math}
           \mathrm{rot}\; \widetilde{\mathbf{H}}
          =
          (\partial \widetilde{\mathbf{D}}/ \partial t)
          +
          \widetilde{\mathbf{j}}
\end{math}),
one can replace the partial time derivative in the first term of the sum (\ref{AbrahamForceStationaryApproximation:a}), which 
may be assigned formally to the "stationary displacement current" or simply to the current density
\begin{math}
          \widetilde{\mathbf{j}},
\end{math}
 by the
\begin{math}
   \mathrm{rot}  \mathbf{\widetilde{H} }
\end{math}
from the first of Eqs (\ref{MaxwellEq:a}), that results in 
\begin{equation}
\label{AbrahamForceStationaryApproximation:b}
\begin{split}
    \left( 1-\frac{1}{\widetilde{\varepsilon}} \right)
    \left[
    \mathrm{rot}  \mathbf{\widetilde{H} }
        \times
    \mathbf{\widetilde{B} }
    \right]
    = \\ =
    \left( 1-\frac{1}{\widetilde{\varepsilon}} \right)
    \left[
    \frac{ k''_{ef} }{ \mu_{0} }  \widetilde{B}_{y}(0) \mathbf{i}
    \times
    \widetilde{B}_{y}(0) \mathbf{j}
    \right].
\end{split}
\end{equation}
This expression 
shows that 
Abraham force
achieves the highest magnitude when the spatial 
inhomogeneity
\begin{math}
    k''_{ef}
\end{math}
gets the largest value. 
It corresponds 
to a half of
the London  penetration depth
\begin{math}
    \lambda_{0}/2
    =
    (2k_{0}'')^{-1}
    =
    (m/\mu_{0}n4e^{2})^{1/2}
\end{math}
 taking into 
account a factor of 2 
that differentiates 
the Eqs (\ref{AbrahamForce2}) and 
(\ref{AbrahamForceStationaryApproximation:b}) in the stationary approximation
\begin{math}
    \omega
    =
    0 .
\end{math}


\section{Formal backgrounds of the electrodynamic nonlocality}
\label{Formal backgrounds} 



To find the origin of this double ratio formal backgrounds of the electrodynamic nonlocality
is
deduced in this section with an emphasis on the analysis of nonself-adjoint problems. Since it may be not quite comprehensible for general interest readers with a background in
the condensed matter
due to the presence of rather abstract technical formulations, this section can be 
skipped in first reading.

The double 
ratio
\begin{math}
    k''_{ef}
    =
    2 k''_{A}
\end{math}
in the force expressions 
can be 
understood
considering
that the operator
\begin{math}
   \mathrm{rot}  \mathbf{\widetilde{H} }
\end{math}
in the Eq.(\ref{AbrahamForceStationaryApproximation:b}) acts for  the "stationary displacement current", 
i. e. for the  displacement-current operator
\begin{math}
         (\partial \widetilde{\mathbf{D}}/ \partial t)
\end{math}
in the zero frequency limit. 
So it should resemble fundamental properties of a displacement current. 
Nonlocality is its basic property that 
comes into definitions of electric 
and magnetic
polarisation
and inductions via the spatially dependent electric 
and magnetic moments. 
In the continuous media electrodynamics \cite{lifshiz2,lifshiz,SilRuh} these moments are introduced as the 
electric 
or magnetic moment formed on 
the boundary surface  of the body 
and related to 
a geometric shape of the surface. 
The boundary  surface 
usually is approximated by the certain 
quadratic form \cite{Shilov} of 
vectors in a 
three-dimensional linear space,
which is a mathematical abstraction of our empirical coordinate space with the real-valued
euclidean
square.
This linear space 
is a linear manifold spanned by the three real spatial basis vectors \cite{Shilov} with a 
basic reference standard of the length 
defined materially. It may be 
considered as the space
\begin{math}
        V_{3}
\end{math}
the elements of 
which are 
vectors (directed line segments) subject to certain suitably defined operations. 
These vectors are studied in three-dimensional analytic geometry and mechanics.
\begin{math}
        V_{3}
\end{math}
is a linear space over the field
\begin{math}
        R
\end{math}
of real numbers \cite{Shilov}. This mathematical abstraction allows
describing
empiric 
physical quantities numerically as 
vector 
or scalar fields in that, given by experiences, real-valued euclidean coordinate space with  three real 
spatial basis vectors. Being defined in this abstract linear space
\begin{math}
        V_{3}
\end{math}
the scalar, vectorial and mixed products of vectors are assumed to be applicable as well to the vectors of the fields of physical quantities as if those belong to the linear space. Furthermore,
in 
differential operators of vector 
or scalar fields of physical quantities, such as Hamilton or Laplace operators, it is these three-dimensional real spatial variables,
concerning
those the spatial derivatives are taken. One can remember 
differential operators in the equations of Maxwell (\ref{MaxwellEq:a}), (\ref{MaxwellEq:b}), 
or of Newton with 
a gradient of a scalar field potential, vector fields of electric magnitudes 
or of velocities, scalar fields of spatial densities 
of 
an electric charge 
or of 
a mass. 
The assumption of applicability of the products of three-dimensional vectors and of the differential field operators to the vector manifold of 
empiric physical fields as if it 
belongs to the 
space
\begin{math}
        V_{3}
\end{math}
allows
using
those in the equations, which explain the physical phenomena.

For this abstract boundary surface to be an electrodynamics boundary surface, i. e. to produce any observable physical effect, it should separate the bounded regions of the space 
by different values of at least the
one 
more real-valued variable parameter, which may vary independently of variations of the three spatial coordinates. 
It corresponds to the inhomogeneous
distribution of this parameter over the space
\begin{math}
        V_{3} .
\end{math}
In Maxwell equations,
such novel independent variables are 
the constitutive parameters or 
a combination of 
these parameters, which
characterise
material properties. 
Those are the specific bulk densities of a charge and of a mass as well as the absolute permittivity, which 
is a combination of these densities and of a frequency, 
i. e. of the time-related parameter. The 
absolute permittivity (cf. \cite{lifshiz2,lifshiz,SilRuh,VolchKarPer,mende} and Eq. (\ref{PermittivityConductivity})) 
\begin{equation*}
\begin{split}
          \widetilde{\varepsilon}_{a}
          &=
          \widetilde{\varepsilon}_{p}
          \varepsilon_{0}
          -
          i \widetilde{\sigma} / \omega
          =
          \widetilde{\varepsilon}_{p}
          \varepsilon_{0}
          -
          i \varepsilon_{0} \omega_{p}^{2} / [\omega( \widetilde{\omega}_{k} + i\omega)]
          = \\
					&=
          \widetilde{\varepsilon}_{p}
          \varepsilon_{0}
          -
          i / [\lambda_{0}^{2}\mu_{0}\omega( \widetilde{\omega}_{k} + i\omega)]
          = \\
					&=
          \widetilde{\varepsilon}_{p}
          \varepsilon_{0}
          -
          i n e^{2}/ [m \omega( \widetilde{\omega}_{k} + i\omega)]
          = \\
					&=
          \widetilde{\varepsilon}_{p}
          \varepsilon_{0}
          -
          i (n e)^{2}/ [(n m) \omega( \widetilde{\omega}_{k} + i\omega)]
\end{split}
\end{equation*}
depends 
on the 
charge
\begin{math}
        ( n e ) 
\end{math}
and
mass
\begin{math}
        ( n m )
\end{math}
densities, 
on the time
\begin{math}
        ( 1 / \omega ) ,
\end{math}
and 
combines 
the basic 
units of length, mass, time and electric charge.

These 
additional real-valued variables of mass, time and 
charge are independent of variations of the three spatial length coordinates 
and consequently should result in 
the 
greater than 3 dimension 
of the abstract linear 
space
over the field
\begin{math}
        R 
\end{math}
of real numbers \cite{Shilov} to proceed 
the description of empiric 
physical fields 
as the vectors 
existing 
in 
a linear space, but with a number of real-valued 
coordinates greater than 3. 
The linear description allows
using
advantages of finite-dimensional linear spaces  \cite{Shilov} given by the application of linear operations in most powerful techniques.

The increased dimension 
requires
an appearance of the respective number of 
additional basis vectors, which are linear-independent of 
three real spatial basis vectors of 
the three-dimensional linear space 
\begin{math}
        V_{3}
\end{math}
comprising a linear manifold spanned by these three real 
basis vectors. Since the given by experiences dimension 
of our coordinate space is limited by three real spatial basis vectors, e. g.
\begin{math}
        \mathbf{i},\,\mathbf{j},\,\mathbf{k} ,
\end{math}
the additional basis vectors must be 
\emph{notional}, not existing in our three-dimensional space, e. g.
\begin{math}
       i \mathbf{i},\,i \mathbf{j},\,i \mathbf{k} 
\end{math}
\cite{Shilov}, because in the linear manifold spanned by these three real 
basis vectors there is no 
vacant space for the fourth or higher linear-independent \emph{real} vector.
The 
variables of mass, time and charge are meant as real-valued so the complex manifold of vectors of physical fields with additional imaginary basis vectors should be considered as
vectors of the real linear space
\begin{math}
        \mathbf{C}_{6}
\end{math}
over the field
\begin{math}
        R 
\end{math}
of real numbers \cite{Shilov} with the real pseudo-euclidean square 
\cite{Rashevskii} that results from the mutual orthogonality 
of the sextet of all real and imaginary basis vectors. The additional basis notional vectors 
\begin{math}
        i\mathbf{i},\,i\mathbf{j},\,i\mathbf{k} 
\end{math}
have in this 
real pseudo-euclidean space the pure imaginary unitary value of the "length" \cite{Rashevskii,Bitsadze}, which is derived from the 
pseudo-euclidean square 
defined 
in the same manner as in the real euclidean 
space
\begin{math}
        V_{3} 
\end{math}
and agrees with 
the requirement of nonexistence of these notional vectors in the 
space
\begin{math}
        V_{3} . 
\end{math}

The similar linear space has been considered formally in more details \cite{Dresvyannikov} when seeking a solution of the nonself-adjoint Laplace operator 
derived from Maxwell equations, which corresponds to the eigenvalue in the form of the complex euclidean square 
of some vector
\begin{math}
           \mathbf{\widetilde{k}}
\end{math}
in the real-base three-dimensional euclidean  space, but generally with  complex components. 
This vector --  the wave vector  may be considered as a vector of the three-dimensional complex euclidean space \cite{Rashevskii}
\begin{math}
        \mathbf{C}_{3}
\end{math}
over the field
\begin{math}
        C 
\end{math}
of complex numbers \cite{Shilov}. The similarly specified complex $n$-dimensional space was introduced in finding a modification of the Jordan matrix suitable for the case of a real space \cite{Shilov}.
In the three-dimensional complex euclidean space the scalar, vectorial and mixed products of vectors may be considered formally, at least
in the 
component form, as defined in the real euclidean 
space
\begin{math}
        V_{3} .
\end{math}
However, the derivatives in differential operators are taken with respect to the real spatial variables \cite{Dresvyannikov}.


Certain general geometric 
restrictions 
have been noted \cite{Dresvyannikov} concerning a behaviour of vectors 
from the complex euclidean space if those are considered in the real 
three-dimensional space, which is associated with the physical reality. The restrictions 
relate 
to a 
known duality \cite{Shilov,Rashevskii} of the representation of the one and the same complex 
affine space as the spaces of different dimension 
over the fields of complex (e. g.
\begin{math}
        \mathbf{C}_{3}
\end{math}
over the field
\begin{math}
        C )
\end{math}
or real (e. g.
\begin{math}
        \mathbf{C}_{6}
\end{math}
over the field
\begin{math}
        R) 
\end{math}
numbers. Subscripts denote here the number of basis vectors. This duality is closely related to the fundamental theorem of algebra, to polynomial factoring with coefficients in the field
\begin{math}
        R 
\end{math}
of real numbers and to the real Jordan canonical form \cite{Shilov}.
In general,
there is no canonical basis in which the matrix of a linear operator acting in a real \emph{n}-dimensional space
\begin{math}
        \mathbf{R}_{n} 
\end{math}
takes the Jordan form \cite{Shilov}, if only because the characteristic polynomial of the operator can have imaginary roots. To find a modification of the Jordan matrix suitable for the case of a real space the linear transformation is used which constructs a basis in the real space
\begin{math}
        \mathbf{R}_{n} 
\end{math}
by replacing each pair of the complex conjugate vectors of the Jordan basis by a pair of real vectors \cite{Shilov}. The transformation resembles the similar one known in theories of superfluidity and superconductivity as 
Bogoliubov transformation.
The different dimensionality of complex and real representations of 
the unique complex 
affine space 
results in different 
possible definitions (see Sections 8.2, 9.2, 9.4 in \cite{Shilov}) 
of the products of vectors 
in these 
complex 
affine spaces, including the mentioned above pseudo- and complex euclidean square \cite{Rashevskii}. 
That 
ambiguity 
could expand \cite{Dresvyannikov} the 
variety of possible  solutions of the nonself-adjoint problems 
and 
might occur to be useful in 
such the first, linear approach to finding
solutions and 
to the future analysis of obtained solutions, which in general should be verified empirically, however.

In the second, nonlinear approach the 
restrictions are a result of a consideration of the vectors from the real linear 
space
\begin{math}
        \mathbf{C}_{6}
\end{math}
over the field
\begin{math}
        R 
\end{math}
of real numbers, 
which is the linear manifold spanned by the sextet of  real and imaginary basis vectors and 
has an even dimensionality as the real number representation of the complex affine space \cite{Shilov,Rashevskii}. 
To the contrary,
we consider such vectors as the vectors of the fields of physical quantities 
as if those 
reveal themselves in the three-dimensional space
\begin{math}
        V_{3} 
\end{math}
over the field
\begin{math}
        R 
\end{math}
of real numbers with the real 
original basis
\begin{math}
        \mathbf{i},\,\mathbf{j},\,\mathbf{k} ,
\end{math}
a dimension 
of which is not even.
As a result,
this three-dimensional real space 
complemented by the complex vector manifold of 
physical fields 
cannot
be in general a three-dimensional linear space due to 
its odd dimensionality. 
Such
a three-dimensional
combination 
of the space and
the
manifold 
cannot
be the real number representation of the complex \emph{affine} space, the dimensionality of which should be even \cite{Shilov,Rashevskii}.
Consequently this combination can not be a subspace 
of the complex space
\begin{math}
        \mathbf{C}_{3}
\end{math}
over the field
\begin{math}
        C 
\end{math}
or of the real space 
\begin{math}
        \mathbf{C}_{6}
\end{math}
over the field
\begin{math}
        R .
\end{math}
It may form only \cite{Dresvyannikov} 
a three-dimensional factor space \cite{Shilov} of the
\begin{math}
        \mathbf{C}_{6}
\end{math}
over the field
\begin{math}
        R 
\end{math}
with respect to the kernel subspace consisting, for example, of the pure imaginary vectors of this 
real space
\begin{math}
        \mathbf{C}_{6} ,
\end{math}
which transforms into the null-vector of the real-base three-dimensional space
\begin{math}
        V_{3} 
\end{math}
over the field
\begin{math}
        R
\end{math}
of real numbers. 

An injection of the space
\begin{math}
        V_{3} 
\end{math}
by the vector manifold of the empiric physical fields 
allows 
to use
the suitably defined and verified empirically operations of products of three-dimensional vectors and 
the differential field operators 
in expressions, resembling empiric physical laws, 
for all vectors of the manifold as if it belongs to the space
\begin{math}
        V_{3} .
\end{math}
However, in this second approach the space
\begin{math}
        V_{3} 
\end{math}
can not continue to be
the affine space 
being combined with the manifold. It means that the "physical space", comprising the combined vector manifolds of the space
\begin{math}
        V_{3} 
\end{math}
and of the empiric physical fields, 
loses 
main attributes of the \emph{linear three-dimensional} 
space over the field
\begin{math}
        R
\end{math}
of real numbers, which 
include the addition rule of vectors and the rule for multiplication of a vector by a number together with respective axioms 
obeyed by these 
operations \cite{Shilov}. 
The vector manifold of empiric physical fields, being combined with 
the space
\begin{math}
        V_{3} , 
\end{math}
transforms it into the 
\emph{nonlinear three-dimensional} vector  manifold corresponding to the three-dimensional factor space \cite{Shilov} of the
\begin{math}
        \mathbf{C}_{6}
\end{math}
over the field
\begin{math}
        R . 
\end{math}
Consequently there is no 
real three-dimensional basis which could produce this manifold as a linear manifold over the field
\begin{math}
        R
\end{math}
of real numbers spanned by three vectors of this real basis. 
It means that such basis could not 
be obtained by any linear 
mapping of 
the 
original real three-dimensional basis
\begin{math}
        \mathbf{i},\,\mathbf{j},\,\mathbf{k} 
\end{math}
of the space
\begin{math}
        V_{3} 
\end{math}
which would be described by an
\begin{math}
        3\times3 
\end{math}
matrix of real constants. 
Otherwise,
the 
existence of 
a linear transformation, which could produce such basis, should lead to the linearity of a vector manifold of the "physical space" due to the general $K$-isomorphism of all $n$-dimensional (here
\begin{math}
        n
        =
        3 )
\end{math}
linear spaces  over a field
\begin{math}
        K
\end{math}
\cite{Shilov}.

Thus 
a linear 
mapping of
\begin{math}
        V_{3} 
\end{math}
should be replaced by a general nonlinear transformation to describe a vector manifold of the "physical space". Such transformation may be thought as an injection 
of the additional vector manifold of physical fields into the linear space
\begin{math}
        V_{3} 
\end{math}
accompanied by the tensile-compressive deformation of the last. It can be described by 
mapping the three-dimensional euclidean linear space
\begin{math}
        V_{3} 
\end{math}
into its three-dimensional "frames" occupied by the combined euclidean vector manifold of this former space
\begin{math}
        V_{3} 
\end{math}
and of physical fields and can be defined, for example, by an explicit real function
\begin{math}
        \mathbf{y}
        =
        \mathbf{y} ( \mathbf{x} ) , 
\end{math}
where
\begin{math}
        \mathbf{x}
        =
        ( x_{1}, x_{2}, x_{3} ) 
\end{math}
and
\begin{math}
        \mathbf{y}
        =
        ( y_{1}, y_{2}, y_{3} ) 
\end{math}
are 
vectors of the space
\begin{math}
        V_{3} 
\end{math}
and of the combined "physical" manifold respectively, by 
a more general implicit relation 
\begin{math}
        F(\mathbf{x}         ,
        \mathbf{y})
        =
        0 ,
\end{math}
or by 
a parametrically definable function.

In accordance with \cite{Bitsadze} let
\begin{math}
        d \mathbf{y}\,
        d \mathbf{y}
\end{math}
and
\begin{math}
        d \mathbf{x}\,
        d \mathbf{x}
\end{math}
be euclidean squares of the "distance" between the points (vectors)
\begin{math}
        \mathbf{y},\,\,
         \mathbf{y}
         +
        d \mathbf{y} 
\end{math}
and
\begin{math}
        \mathbf{x},\,\,
         \mathbf{x}
         +
         d \mathbf{x} .
\end{math}
Those are the comparable line elements in the nonlinear vector manifold 
and in the linear space
\begin{math}
        V_{3} , 
\end{math}
from which this 
nonlinear manifold has been transformed. 
Let
\begin{math}
        d \mathbf{\sigma}^{2}
\end{math}
and
\begin{math}
        d \mathbf{s}^{2} 
\end{math}
denote the euclidean squares, 
respectively. The mapping is Gauss conformal 
if there is a positive scalar function
\begin{math}
        \lambda( \mathbf{x} )
\end{math}
that satisfies the relation
\begin{math}
        d \mathbf{\sigma}^{2}
         =
        \lambda( \mathbf{x} )
        d \mathbf{s}^{2} .
\end{math}
It is angle-preserving and Liouville's theorem on conformal mappings in 
euclidean space states \cite{Bitsadze} that any smooth conformal mapping on a domain of
\begin{math}
        \mathbf{R}_{n},
\end{math}
where
\begin{math}
        n
        >
        2 ,
\end{math}
can be expressed as a composition of a finite number of translations, similarities, orthogonal transformations and inversions: they are M\"{o}bius transformations (in
\begin{math}
        n
\end{math}
dimensions). This theorem severely limits the variety of possible conformal mappings in
\begin{math}
        \mathbf{R}_{3}
\end{math}
(here in
\begin{math}
        V_{3}) ,
\end{math}
and higher-dimensional spaces.

Three of four possible elementary operations of translations
\begin{equation} 
        \mathbf{y}
        =
        \mathbf{x}
        +
        \mathbf{h}\, ,
          \label{Translations}
\end{equation}
similarities
\begin{math}
        \mathbf{y}
        =
        \mu
        \mathbf{x} ,
\end{math}
orthogonal transformations
\begin{math}
        \mathbf{y}
        =
        C
        \mathbf{x} ,
\end{math}
where
\begin{math}
        \mathbf{h}
        =
        ( h_{1}, h_{2}, h_{3} ) 
\end{math}
is the constant vector,
\begin{math}
        \mu
\end{math}
- the real constant scalar, and
\begin{math}
        C
\end{math}
- the orthogonal matrix \cite{Shilov,Bitsadze} of real constants, are the linear maps 
with
\begin{math}
        \lambda
        =
        1
\end{math}
for the first and third operations and with
\begin{math}
        \lambda
        =
        \mu^{2} 
\end{math}
for the 
second operation \cite{Bitsadze}. To 
ensure the inevitable nonlinearity of a vector manifold of the "physical space" these conformal maps 
of translations, similarities, orthogonal transformations or any composition comprising only these elementary linear maps should be 
eliminated from possible general 
transformations of 
\begin{math}
        V_{3} 
\end{math}
into the "physical space". Conformal maps 
of translations and similarities 
and of orthogonal transformations 
may be considered, respectively, as global translational (translational-dilatational, in general) and rotational symmetries of 
a finite-dimensional affine space. 
It means that these global symmetries correspond to linear maps of the affine space into itself. 
Consequently,
the physical nonlocality, 
which is embedded in Maxwell equations and has been discussed in the beginning of this section,
leads to 
the inevitable nonlinearity of the vector manifold of the non-empty "physical space" resulting in the broken global translational and rotational symmetries of the 
formerly empty affine space
\begin{math}
        V_{3} .
\end{math}
These global symmetries may be considered as 
necessary and sufficient
conditions of the linearity of a real finite-dimensional vector space (in
\begin{math}
        n
\end{math}
dimensions,
where
\begin{math}
        n
        >
        2 ),
\end{math}
a failure of which corresponds reasonably to the 
supposed above 
inhomogeneous distribution of 
constitutive parameters over the space 
\begin{math}
        V_{3} .
\end{math}
Necessity follows from the definition \cite{Shilov} of linear space and sufficiency has been proven just above.

Thus the only allowed one of four possible in
\begin{math}
        V_{3} 
\end{math}
elementary operations of global Gauss conformal maps 
is the 
inversion 
\begin{equation} 
        \mathbf{y}
        =
        \frac{\mathbf{x}}{(\mathbf{x}\cdot\mathbf{x})}\,\,
          \label{Inversions}
\end{equation}
of the affine space
\begin{math}
        V_{3} 
\end{math}
into a nonlinear
vector manifold of the "physical space" with the nonlinear
scalar function
\begin{math}
        \lambda( \mathbf{x} )
        =
       (\mathbf{x}\cdot\mathbf{x})^{-2}
\end{math}
\cite{Bitsadze}, where
\begin{math}
         (\mathbf{x}\cdot\mathbf{x})
\end{math}
is the scalar product. 
Any arbitrary composition of elementary conformal maps of the space
\begin{math}
        V_{3} 
\end{math}
into the nonlinear manifold of "physical space", which may be required 
in the solution of an electrodynamic problem, should necessarily incorporate the 
nonlinear
map of inversion among four possible types of conformal maps.
This conclusion illustrates the mentioned above
tensile-compressive deformation of the "physical space". It is a consequence of the 
duality in
the
representation
of the complex affine space as the complex or real space and finds good agreement with the conception of the phase space.


The ambiguity in possible definitions of 
products of vectors in finite-dimensional complex 
affine spaces can expand, 
as it has been previously discussed in this section, the 
variety of possible  solutions of the
nonself-adjoint
problems, if only because the 
adjoint operators 
are defined
concerning
the 
fixed bilinear form \cite{Shilov,Rashevskii}. This ambiguity can lead also to different
nonlinear metric properties of 
a vector manifold of the "physical space". It may be illustrated 
by the simplest 
one-dimensional complex space
\begin{math}
        \mathbf{C}_{1} 
\end{math}
and by its two-dimensional real number representation
\begin{math}
        \mathbf{C}_{2} .
\end{math}
Let
\begin{math}
        d \mathbf{z}\,
        d \mathbf{z}
\end{math}
be again a scalar square of the "distance" between the complex points (complex vectors)
\begin{math}
        \mathbf{z},\,\,
         \mathbf{z}
         +
        d \mathbf{z} ,
\end{math}
i. e. of the line element, where
\begin{math}
        \mathbf{z}
        =
        ( x 
        +
        i
        y )
        \mathbf{i}
\end{math}
and
\begin{math}
        \mathbf{z}
         +
        d \mathbf{z}
        =
        (( x 
        +
        d x ) 
        +
        i
        (y
        +
        d y ) )
        \mathbf{i}
\end{math}
are 
vectors of the complex space
\begin{math}
        \mathbf{C}_{1} .
\end{math}
Let
\begin{math}
        \mathbf{C}_{1} 
\end{math}
be a 
complex
unitary space 
with the hermitian scalar 
product \cite{Shilov} of the line element
\begin{math}
        d \mathbf{z}\,
        d \mathbf{z}
        =
        (( 
        d x 
        +
        i
        d y ) \,
        \mathbf{i},
        \,
        (
        d x 
        - 
        i
        d y ) \,
        \mathbf{i} )
        =
        d x^{2}
        +
        d y^{2} 
\end{math}
or, alternatively,
be a 
complex 
euclidean space \cite{Rashevskii} with the euclidean scalar square 
\begin{math}
        d \mathbf{z}\,
        d \mathbf{z}
        =
        (( 
        d x 
        +
        i
        d y ) \,
        \mathbf{i},
        \,
        (
        d x 
        +
        i
        d y ) \,
        \mathbf{i} )
        =
        ( d x^{2}
        -
        d y^{2} )
        +
        2 \,i \, dx \,
        dy .
\end{math}
Consider the respective real space
\begin{math}
        \mathbf{C}_{2} 
\end{math}
consisting of the formal sums
\begin{math}
        \mathbf{z}
        =
        x \,\,
        \mathbf{i}
        +
        y
        (i\,\,
        \mathbf{i})
\end{math}
and
\begin{math}
        \mathbf{z}
         +
        d \mathbf{z}
        =
        ( x 
        +
        d x ) \,\,
        \mathbf{i} 
        +
        (y
        +
        d y )
        ( i \,\,
        \mathbf{i}) ,
\end{math}
which are 
vectors in a complex orthonormal basis of the
real space
\begin{math}
        \mathbf{C}_{2} 
\end{math}
\cite{Shilov} 
(taking account of the pure imaginary unitary "length" of the basis notional vector). 
The following natural scalar products 
may be defined arbitrarily, being not predetermined a priori as well as for the space
\begin{math}
        \mathbf{C}_{1} , 
\end{math}
by the formulae
\begin{math}
        d \mathbf{z}\,
        d \mathbf{z}
        =
        (( 
        d x \, \,\mathbf{i} 
        +
        d y 
        (i\,\mathbf{i})),
              ( 
        d x \, \,\mathbf{i} 
        +
        d y 
        (-i\,\mathbf{i})) )
        =
        d x^{2}
        +
        d y^{2} 
\end{math}
corresponding to the real euclidean space with the hermitian scalar product 
of the basis notional vector, 
or alternatively,
by
\begin{math}
        d \mathbf{z}\,
        d \mathbf{z}
        =
        (( 
        d x \, \,\mathbf{i} 
        +
        d y 
        (i\,\mathbf{i})),
              ( 
        d x \, \,\mathbf{i} 
        +
        d y 
        (i\,\mathbf{i})) )
        =
        d x^{2}
        - 
        d y^{2} 
\end{math}
corresponding to the real
pseudo-euclidean space \cite{Rashevskii} with the 
euclidean scalar square of the basis notional vector. 

A general nonlinear transformation into 
a vector manifold of the "physical space" 
for the case of 
\begin{math}
         \mathbf{C}_{2} 
\end{math}
can be described by 
mapping the 
one-dimensional euclidean linear space
\begin{math}
        V_{1} 
\end{math}
into its 
one-dimensional "frames" occupied by the combined 
vector manifold of this former space
\begin{math}
        V_{1} 
\end{math}
and of physical fields and can be defined, for example, by 
a parametrically definable function using the component
\begin{math}
        x
\end{math}
of the real basis vector as a parameter, i. e.
\begin{math}
        \mathbf{z}(x)
        =
        x \,\,
        \mathbf{i}
        +
        y(x)
        (i\,\,
        \mathbf{i}) .
\end{math}
It allows to 
estimate a respective transformation of a scalar square of the line element. Thus for considered above scalar products of the line element
\begin{math}
        d \mathbf{\sigma}^{2}
\end{math}
one obtains the following
$x$-dependencies
\begin{math}
        d \mathbf{\sigma}^{2}
         =
        \lambda( x ) 
        d \mathbf{s}^{2} :
\end{math}
\begin{math}
        [
        1
        +
        (dy/dx)^{2} 
        ]
        d x^{2} 
\end{math}
for a complex unitary space
\begin{math}
         \mathbf{C}_{1} 
\end{math}
with the hermitian product 
as well as for a 
real euclidean space
\begin{math}
         \mathbf{C}_{2} 
\end{math}
with the hermitian product of the basis notional vector,
\begin{math}
        [
        1
        +
        i(dy/dx)
        ]^{2} 
        d x^{2}
        =
        [
        1
        -
        (dy/dx)^{2} 
        +
        2 i
        (dy/dx)
        ]d x^{2} 
\end{math}
for a complex 
euclidean space \cite{Rashevskii}
\begin{math}
         \mathbf{C}_{1} 
\end{math}
with the 
euclidean scalar square,
and
\begin{math}
        [
        1
        -
        (dy/dx)^{2} 
        ]d x^{2} 
\end{math}
for a 
real pseudo-euclidean space \cite{Rashevskii}
\begin{math}
         \mathbf{C}_{2} 
\end{math}
with the 
euclidean scalar square of the basis notional vector. 
A scalar function
\begin{math}
        \lambda( x ) 
\end{math}
characterising
a transformation of the metric is generally the nonlinear square-law real function, 
except the complex function for the 
complex 
euclidean space 
\begin{math}
         \mathbf{C}_{1} .
\end{math}
This transformation function is of an elliptic type for the spaces
\begin{math}
         \mathbf{C}_{1} 
\end{math}
and
\begin{math}
         \mathbf{C}_{2} 
\end{math}
with the hermitian product, and of a 
hyperbolic type for the spaces
\begin{math}
         \mathbf{C}_{1} 
\end{math}
(at least for the real part of
\begin{math}
        \lambda( x ) 
        )  
\end{math}
and
\begin{math}
         \mathbf{C}_{2} 
\end{math}
with the euclidean scalar square. 
In this last
case 
of the pseudo-euclidean 
space
\begin{math}
         \mathbf{C}_{2} 
\end{math}
there is a limitation 
of the value of 
derivative in the inequation
\begin{math}
        \lambda
        ( x )
        =
        [
        1
        -
        (dy/dx)^{2} 
        ]
        \geq
        0
\end{math}
to provide the 
real, not pure imaginary 
length of the line element 
for it would not belong to the zero class of a factor space and could 
be measurable in the "physical space" of real vectors.


\section{Effective parameters and Abraham force} 
\label{Effective constitutive parameters} 

The nonlocality may be 
considered as an indication 
of 
inhomogeneity of the "physical space" 
corresponding to the broken global translational symmetry, that 
has been 
formally demonstrated in 
the previous section \ref{Formal backgrounds}.
In the problem considered here the nonlocality is revealed as a requirement of electroneutrality, i. e. as a requirement of an absence of extrinsic charges (see the first of Eqs (\ref{MaxwellEq:b})). 
\subsection{Formal effects of electroneutrality} 
\label{Outcomes of the electroneutrality conditions}
The requirement of electroneutrality results in the zero values of the bulk integral
electric charge and of the total kinetic (true) momentum
\begin{math}
   \textit{\textbf{p}} 
   =
   0 
\end{math}
in the absence of extrinsic currents, if 
the total momentum can be written in the form characteristic of the rigid medium
\begin{math}
   \textit{\textbf{p}} 
   =
   m \textit{\textbf{v}} 
\end{math}
where the velocity
\begin{math}
   \textit{\textbf{v}} 
\end{math}
is
the same constant vector for all particles or parts of the rigid medium. Here the velocity
\begin{math}
   \textit{\textbf{v}} 
\end{math}
of an intrinsic charge carrier is measured relative to the 
rest reference coordinate system which is exterior to the conductor,
\begin{math}
    m
\end{math}
is the parameter expressed in units of mass. The approximation of a rigid medium correlates with the discussed in section \ref{Superconductors and perfect conductors}
feasibility of "acoustic" conditions 
for the spatially dispersive plasmon 
polariton frequency and with Landau criterion for superfluidity.
The presence of the critical minimal velocity for intrinsic elementary excitations in the system leads to the movement of the system as a whole, i. e. as a rigid medium when the velocity of the movement is lower than the critical velocity for intrinsic excitations.

The latter requirement of the zero 
total 
momentum just corresponds 
to 
inhomogeneity of the space relative to the movement of participate matter 
considered as a rigid medium. 
Otherwise,
the 
constant non-zero 
vector of the total true (kinetic) momentum
\begin{math}
   \textit{\textbf{p}} 
      =
   m \textit{\textbf{v}}
   \neq
   0 
\end{math}
would correspond to the invariant state of the "physical space" comprising current carriers. 
The instant states of 
such physical system 
could only differ 
in just spatial coordinates of current carriers. 
These variations of coordinates would not 
affect all other parameters of the momentum-conserved state. 
Thus 
coordinates of current carriers in every 
instant state 
of such physical system could be 
linearly transformed 
by translations of the space
\begin{math}
        V_{3},
\end{math}
in which the coordinates should be determined. 
The 
constant vector in  the linear map of translations 
(\ref{Translations}) would be equal to 
\begin{math}
        \mathbf{h}
        =
         \textit{\textbf{p}} \Delta t
         /
         m 
   =
   \textit{\textbf{v}}  \Delta t ,
\end{math}
where
\begin{math}
     \Delta t
\end{math}
should be the 
respective interval of time.
However,
the 
linear conformal map of translations
(cf. Eq. (\ref{Translations})) 
must be excluded from possible general transformations
of a finite-dimensional affine space
\begin{math}
        V_{3} 
\end{math}
into the nonlinear vector manifold of the "physical space" 
as it has been formally 
derived in section \ref{Formal backgrounds}.
Consequently,
the global translational symmetry of a finite-dimensional affine space should be broken in the nonlinear real vector manifold of the "physical space" where coordinates of current carriers 
would be determined. 
The non-zero 
vector of the total true (kinetic) momentum
\begin{math}
   \textit{\textbf{p}} 
   \neq
   0 
\end{math}
can not 
belong to this 
"coordinate physical space" in 
the approximation of rigid medium. 
Formal backgrounds of the previous section 
allow
concluding
that non-zero 
vectors of the total true (kinetic) momentum
can belong only to the zero class of a factor space, have the pure imaginary "length" and 
cannot
be measurable 
in the real "coordinate physical space" in this approximation. 

These inferences
cannot
be 
generalised
with respect
the composite non-rigid medium 
of current carriers even under conditions of electroneutrality since the mass factor, which 
has been additive in the rigid-medium approximation
\begin{math}
    m
   =
         \sum m_{i} ,
\end{math}
comes in the linear combination with possibly different vectors of velocity and
cannot
be involved reciprocally in the vector of translation
\begin{math}
        \mathbf{h} .
\end{math}
As a result the map of translations (cf. Eq. (\ref{Translations})) 
for the composite non-rigid medium 
fails to be linear on the 
kinetic 
momentum
\begin{math}
        (\mathbf{h} 
        \sim
         \textit{\textbf{p}}) 
\end{math}
in general.
The 
different behaviour of rigid and composite non-rigid media correlates with earlier discussions of electrodynamic nonlocality (cf. \cite{TowardsQTL} and references therein).

This conclusion explains also why the superconducting effects, caused by the absence of the dc dissipation, for example the Meissner effect, demonstrate the empirically observed parameters 
described by the solutions 
determined by 
conditions of the spatial force resonances with the pure imaginary value of 
the expression (\ref{ComponentOfMaxwellianFieldStressTensor}) for the spatial density of the momentum flux \cite{Dresvyannikov,KarLT26}.  
The pure imaginary value of the non-zero 
vector of the total true (kinetic) momentum 
is compatible with the approximation of rigid medium. 
On the contrary, the solutions with the pure 
real value of the spatial density of the momentum flux \cite{Dresvyannikov,KarLT26}, which is not compatible with the approximation of rigid medium corresponding to superconducting properties, describe the normal state effects, for example the anomalous skin-effect.


A displacement current is 
a movement of the intrinsic charge carriers and can possess only
the
quasi-momentum
\begin{math}
   \hbar
   \textit{\textbf{k}} 
   \neq
   0 
\end{math}
in the rigid-medium approximation. The
quasi-momentum
vector manifold corresponds to the allowed conformal map of inversions (cf. Eq. (\ref{Inversions}) and discussions in section \ref{Formal backgrounds}).

\subsection{Effective parameters} 
\label{Effective parameters}

A stationary current density
\begin{math}
          \widetilde{\mathbf{j}}
\end{math},
substituted into the Eq. (\ref{AbrahamForceStationaryApproximation:b}) instead of the
\begin{math}
         (\partial \widetilde{\mathbf{D}}/ \partial t) 
\end{math}
in the form of
\begin{math}
   \mathrm{rot}  \mathbf{\widetilde{H} } ,
\end{math}
may not satisfy the 
basic requirement of electroneutrality being generally composed of the induced intrinsic and extrinsic components and can have the total kinetic (true) momentum
\begin{math}
   \textit{\textbf{p}} 
   \neq
   0 
\end{math}
even in the rigid-medium approximation.

To
fulfil
the electroneutrality requirement unambiguously one 
can introduce it, together with the condition 
\begin{math}
   \textit{\textbf{p}} 
   =
   0 ,
\end{math}
in the constitutive parameter
\begin{math}
    k''_{ef}
\end{math}
in the Eq.(\ref{AbrahamForceStationaryApproximation:b}) similar to the manner used in the quasiparticle formalism. 
A displacement current alters an electric 
dipole moment, which is defined as an electric 
charge 
multiplied 
by a distance
\begin{math}
    q \cdot  \Delta x .
\end{math}
Let an increment 
of the electric 
dipole moment in the first 
interior 
coordinate system, acquired by a moving charge 
in a time interval
\begin{math}
    t_{0} ,
\end{math}
is equal to
\begin{math}
    q_{m} 
    \Delta x_{m} 
\end{math}
where
\begin{math}
    \Delta x_{m} 
    =
    v_{m} t_{0} 
\end{math}
and 
\begin{math}
    v_{m} 
\end{math}
is the velocity of the charge movement 
in 
this first 
coordinate system, relative to the $x$-axis of which a charge is moving.
Since in 
the surface impedance (plane wave) approximation the three-dimensional electrodynamics problem is reduced to the one-dimensional partial differential equation
\cite{lifshiz2,lifshiz,SilRuh,KarLT26,mende}, 
one can 
use the specific linear density of the mobile charges
\begin{math}
    \eta 
\end{math}
in place of the specific bulk charge density generally  considered \cite{lifshiz2}. 
In
a
conductor,
it should 
be compensated by the same value of
the
specific
linear density of the charge of
the
opposite sign
\begin{math}
    \rho 
\end{math}
to
fulfil
the electroneutrality requirement. It may be treated as 
a charge of the background 
and can be assigned to the 
interior 
coordinate system. In this paradigm 
the moving intrinsic charge 
may be assumed as an integrator of the incoming specific linear density of 
mobile charges in the process of a 
charge motion along the $x$-coordinate, e. g.,
\begin{math}
    q
    =
    \eta \cdot | \Delta x |.
\end{math}
The dipole moment, acquired by a charge 
moving 
relative to the first interior 
coordinate system 
\begin{math}
    q_{m} 
    \Delta x_{m} 
\end{math}
where
\begin{math}
    \Delta x_{m} 
    =
    v_{m} t_{0} , 
\end{math}
due to an absence of extrinsic charges
should be equal to the dipole moment acquired by a charge in 
any interior coordinate system as well as in the second 
interior coordinate system, 
relative to which the charge 
looks like immobile 
in the sense of the zero kinetic (true) momentum 
in the 
exterior rest coordinates,
\begin{math}
    q_{L}
    \Delta x_{L}
\end{math}
where
\begin{math}
    \Delta x_{L}
    =
    v_{L} t_{0}
\end{math}
and
\begin{math}
    v_{L} 
\end{math}
is the velocity of the charge movement in this second 
interior coordinate system. Here the subscript \emph{L}
stands for Langmuir and means that 
the parameters
of this looking-like-in-the-rest charge carrier enter the plasma frequency (\ref{LangmuirPlasmaFrequency}), (\ref{LondonPenetrationDepth}). 
That allows to
fulfil
the requirement of zero kinetic (true) momentum. 

The boundary conditions in directions lateral to the boundary plane are not fixed at any determinate spatial points 
for the 
surface impedance (plane wave) approximation, so the  electroneutrality requirement should result in the local implementation of this requirement for the moving intrinsic charge. 
To assure 
it simultaneously with the
above-considered
condition of zero kinetic (true) momentum and regardless of lateral boundary conditions let us compare the electric dipole moments in both the first and the second interior coordinate systems, which have been introduced here. As it was mentioned previously, 
the electric 
dipole moments 
acquired by a 
charge 
in a time interval
\begin{math}
    t_{0} 
\end{math}
should be 
the same in both interior coordinate systems due to an absence of extrinsic charges
\begin{math}
    q_{m} 
    \Delta x_{m} 
    =
    q_{L}
    \Delta x_{L} . 
\end{math}
Here
\begin{math}
    \Delta x_{m} 
    =
    v_{m} t_{0} , 
\end{math}
\begin{math}
    \Delta x_{L}
    =
    v_{L} t_{0} ,
\end{math}
\begin{math}
    v_{m} 
\end{math}
and
\begin{math}
    v_{L} 
\end{math}
are the respective velocities of the charge movement in the
first interior 
coordinate system, relative to which a charge is moving, 
and in the second 
interior coordinate system, 
relative to which the charge 
looks like immobile in the 
exterior rest reference coordinates. 
The electric charge of the immobile 
carrier
\begin{math}
    q_{L}
\end{math}
corresponds to the charge value entering the Langmuir plasma frequency (\ref{LangmuirPlasmaFrequency}), (\ref{LondonPenetrationDepth}).

The condition of local neutrality for the moving intrinsic charge, which acts as a discussed above integrator of the incoming specific linear density
\begin{math}
    \eta , 
\end{math}
may be formulated as a predicate of the local even parity of a spatial distribution of the bare linear density
\begin{math}
    \rho 
\end{math}
of the background charge of opposite sign
in a vicinity 
of the location point of moving charge. Such a local symmetry of the background
\begin{math}
    \rho 
\end{math}
regions around the peak
\begin{math}
    \eta 
\end{math}
acquired by the charge
\begin{math}
    q_{m}
\end{math}
moving over 
these symmetric 
sections of the
bare, noncompensated background allows
remaining
the global electroneutrality
irrespectively of lateral boundary conditions. 
The desired symmetry of the
bare 
background 
sections can be assured by a movement of the mobile charge with the velocity
\begin{math}
    v_{m} 
\end{math}
relative to the first coordinate system, which itself should move in the opposite direction with the value of velocity
\begin{math}
     v_{L}
     =
     2 v_{m} 
\end{math}
relative to the exterior rest coordinates. A moving intrinsic charge
\begin{math}
    q_{m}
\end{math}
would integrate the incoming specific linear density of 
mobile charges 
\begin{math}
    \eta 
\end{math}
stripping it off the both sides around this moving 
charge. The section of bare background on the one side 
would be formed 
due to the charge
movement 
together with the first coordinate system with the velocity value
\begin{math}
     v_{L}
     =
     2 v_{m} . 
\end{math}
The same section on the opposite side 
would be formed due to the movement of the charge itself relative to the first coordinate system with the velocity value
\begin{math}
    v_{m} . 
\end{math}
Substituting this ratio of velocities in the equality of dipole moments in both intrinsic coordinate systems
\begin{math}
    q_{m} 
    \Delta x_{m} 
    =
    q_{L}
    \Delta x_{L} 
\end{math}
in the form of
\begin{math}
    \Delta x_{m} 
    =
    v_{m} t_{0} , 
\end{math}
\begin{math}
    \Delta x_{L}
    =
    v_{L} t_{0} 
\end{math}
one obtains the relations
\begin{math}
    q_{m} 
    \cdot v_{m} t_{0}
    =
    q_{L}
    \cdot v_{L} t_{0}
    =
    q_{L} \cdot
    2 v_{m} t_{0} ,    
\end{math}
i. e. the double ratio
\begin{math}
    q_{m} 
    =
    2 q_{L}
\end{math}
for the charges in the first and second intrinsic coordinate systems. Since the charge
\begin{math}
    q_{L}
\end{math}
enters 
the Langmuir plasma frequency expressions (\ref{LangmuirPlasmaFrequency}), (\ref{LondonPenetrationDepth}), the charge
\begin{math}
    q_{m}
    =
    2 q_{L}
\end{math}
being substituted in these expressions gives the double ratio for penetration depths and wave numbers
\begin{math}
    k''_{ef}
    =
    2 k''_{A}
\end{math}
in the force expressions 
(\ref{AbrahamForce2}) and 
(\ref{AbrahamForceStationaryApproximation:b}).
So it may be argued that at least two coordinate systems 
should be introduced in an electrodynamics problem to describe the nonlocality properly. 
It would 
be analogous with the situation 
of two travelling waves, which could compose one standing wave, and vice versa.

The double ratio
\begin{math}
    k''_{ef}
    =
    2 k''_{A}
\end{math}
in the force expressions (\ref{AbrahamForce2}) and 
(\ref{AbrahamForceStationaryApproximation:b})
results from the 
non-dissipating direct 
current, 
which provides 
the Meissner effect and 
may be 
presented as the superposition of plasmon polaritons with positive and negative frequencies, corresponding to the zero total frequency, and with equal wavenumbers
\begin{math}
    k_{0}'',
\end{math}
corresponding to the double total wavenumber
\begin{math}
    2k_{0}''
\end{math}
(cf. Eqs (\ref{StaticCurrentDensity}) and (\ref{CurrentDensity})).
The comprehensible generally accepted 
scenario 
emerges from the given explanations, 
according to which the Meissner effect in superconductors is the result of an incomplete cancellation of the diamagnetic and the paramagnetic currents in response to an external magnetic induction field
\begin{math}
    \mathbf{B}
\end{math}
at temperatures below the transition temperature.

Such "pairing" effect should be inherent both for superconductors and for
"normal" perfect
conductors, which may be
non-superconducting
in the
bulk samples of the macroscopic scale even for
the electronic liquid in the collision free limit. This conclusion is supported by the experimental data \cite{Sharvin} on the effective pairing for mesoscopic samples of non-superconducting magnesium
and probably \cite{ICPS98} for nonequilibrium
microscopic
electron-hole droplets in germanium, which expel the static field.
Those can be tentatively related to the effects of the limited system boundary sizes in the spatial dispersion due to the competition of different spatial-type force resonances.



\section{London equations and Abraham force} 
\label{London equations and Abraham force} 

The partial derivative of 
magnetic induction with respect to
time 
in the second of Maxwell equations (\ref{MaxwellEq:a}) may be represented as 
\begin{equation}
\label{MaxwellEq:induc-curr}
\begin{split}
          (\partial \widetilde{\mathbf{B}}/ \partial t)
          &=
          - \mathrm{rot}\; \widetilde{\mathbf{E}}
          = 
          - \mathrm{rot}\; (\widetilde{\mathbf{j}}/ \widetilde{\sigma}) 
          = \\
					&=
          - (i \omega \mu_{0}/(k''_{0})^{2}) \mathrm{rot}\; \widetilde{\mathbf{j}} , 
\end{split}
\end{equation}
where the first London equation \cite{SuperfluidsI,VolchKarPer,mende}
\begin{equation*}
\begin{split}
          d \widetilde{\mathbf{j}}\,(z,t)/ d t 
          &=
          d (\widetilde{\mathbf{j}}\,(z)e^{i \omega t})/ d t 
          =
          i \omega \widetilde{\mathbf{j}}\,(z,t)
          = \\
					&=
          (1/\mu_{0}\lambda_{0}^{2}) 
          \widetilde{\mathbf{E}}
          =
          ((k''_{0})^{2}/\mu_{0}) 
          \widetilde{\mathbf{E}} 
\end{split}
\end{equation*}
has been used to derive the conductivity
\begin{math}
 \widetilde{\sigma}
\end{math}
from Ohm's law 
\begin{math}
          \widetilde{\mathbf{j}}
          =
           \widetilde{\sigma}\widetilde{\mathbf{E}}
          =
          (1/ i \omega \mu_{0}\lambda_{0}^{2})
          \widetilde{\mathbf{E}}
          =
          ((k''_{0})^{2}/ i \omega \mu_{0}) 
          \widetilde{\mathbf{E}} .
\end{math} 
Substituting the Eq. (\ref{MaxwellEq:induc-curr}) and second London equation \cite{SuperfluidsI,mende} 
\begin{math}
           \mathrm{rot}\; \widetilde{\mathbf{j}} 
          =
          -
          (1 / \lambda_{0}^{2})
          \widetilde{\mathbf{H}}
           =
          -
          (k''_{0})^{2} 
          \widetilde{\mathbf{H}}
\end{math} 
into the 
last term of 
Abraham force (\ref{AbrahamForceStationaryApproximation:a}), 
one gets the reinforced expression instead of the Eq. (\ref{AbrahamForceStationaryApproximation:b})
\begin{gather}		
    \left( 1-\frac{1}{\widetilde{\varepsilon}} \right)
    \left[
    \frac{
    \partial \widetilde{\mathbf{D}}
    }{
    \partial t
    }
    \times
    \widetilde{\mathbf{B}}
    \right]
    +
     \left( 1-\frac{1}{\widetilde{\varepsilon}} \right)
    \left[
    \widetilde{\mathbf{D}}
    \times
    \frac{
    \partial \widetilde{\mathbf{B}}
    }{
    \partial t
    }
    \right] 
    = \nonumber
    \\
    =
    \left( 1-\frac{1}{\widetilde{\varepsilon}} \right)
    \left[
    \mathrm{rot} \, \mathbf{\widetilde{H} }
        \times
    \mathbf{\widetilde{B} }
    \right]
    -
    \left( 1-\frac{1}{\widetilde{\varepsilon}} \right)
    \left[
    \widetilde{\mathbf{D}}
    \times
    \mathrm{rot} \, \mathbf{\widetilde{E} }
    \right] 
    = \nonumber
    \\
    =
    \left( 1-\frac{1}{\widetilde{\varepsilon}} \right)
    \left[
    \mathrm{rot} \, \mathbf{\widetilde{H} }
        \times
    \mathbf{\widetilde{B} }
    \right]
    -
    \left( 1-\frac{1}{\widetilde{\varepsilon}} \right)
    \left[
    \widetilde{\mathbf{D}}
    \times
    \frac{i \omega \mu_{0}}{(k''_{0})^{2}}
    \, \mathrm{rot}\; \widetilde{\mathbf{j}} \,
    \right] 
    = \nonumber
    \\
    =
     \left( 1-\frac{1}{\widetilde{\varepsilon}} \right)
    \left[
    \mathrm{rot} \, \mathbf{\widetilde{H} }
        \times
    \mathbf{\widetilde{B} }
    \right]
    -
    \left( 1-\frac{1}{\widetilde{\varepsilon}} \right)
    \left[
    \widetilde{Z}\,
    \widetilde{\mathbf{D}}
    \times
    \frac{1}{k''_{0}}
    \, \mathrm{rot}\; \widetilde{\mathbf{j}} \,
    \right] 
    =
    \nonumber
    \\
    =
    \left( 1-\frac{1}{\widetilde{\varepsilon}} \right)
    \left[
    \mathrm{rot} \, \mathbf{\widetilde{H} }
        \times
    \mathbf{\widetilde{B} }
    \right]
    +
    \left( 1-\frac{1}{\widetilde{\varepsilon}} \right)
    \left[
    \widetilde{B}_{y}(0) \mathbf{i}
    \times
    k''_{0} 
    \, 
    \widetilde{\mathbf{H}} \, 
    \right] 
    =
		\nonumber
    \\
    =
    \left( 1-\frac{1}{\widetilde{\varepsilon}} \right)
    \mu_{0}
    \left[
    \mathrm{rot} \, \mathbf{\widetilde{H} }
        \times
    \mathbf{\widetilde{H} } 
    \right]
    +
		\nonumber
		\\
		+
    \left( 1-\frac{1}{\widetilde{\varepsilon}} \right)
    \left[
    \frac{ k''_{0} }{ \mu_{0} }  \widetilde{B}_{y}(0) \mathbf{i}
    \times
    \widetilde{B}_{y}(0) \mathbf{j}
    \right]
    =
		\nonumber
    \\
    =
    2 
    \left( 1-\frac{1}{\widetilde{\varepsilon}} \right)
    \left[
    \frac{ k''_{0} }{ \mu_{0} }  \widetilde{B}_{y}(0) \mathbf{i}
    \times
    \widetilde{B}_{y}(0) \mathbf{j}
    \right]. \label{2AbrahamForceStationaryApproximation:a}
\end{gather}
Here the modulus (Eq. (\ref{SurfaceImpedanceOnFrequency})) and argument
\begin{math}
          \psi_{1}=\pi/2
\end{math}
of the complex surface impedance
\begin{math}
     \widetilde{Z}
\end{math}
have been used similarly to the Eq. (\ref{AbrahamForce2}). Comparisons of 
the Eqs (\ref{AbrahamForce2}), (\ref{AbrahamForceStationaryApproximation:b}) and (\ref{2AbrahamForceStationaryApproximation:a}) 
reveal the double 
ratios
\begin{math}
    k''_{ef}
    =
    2 k''_{A}
    =
    2 k''_{0}
\end{math}
in the force expressions. 


The equal values 
of penetration depths and wave numbers
\begin{math}
    k''_{0}
    =
    k''_{A}
\end{math}
in the force expressions 
(\ref{AbrahamForce2}) and 
(\ref{2AbrahamForceStationaryApproximation:a}) show that 
the Abraham force
in the Eq. (\ref{AbrahamForce2}) in the stationary limit
\begin{math}
    \omega\rightarrow0
\end{math}
correctly 
deals with the role of the gauge degrees of freedom to define the fields
\begin{math}
    \mathbf{H}
\end{math}
and
\begin{math}
    \mathbf{D} .
\end{math}
The fixed argument and modulus of a complex surface impedance
\begin{math}
          \widetilde{Z}
\end{math}
determined by 
conditions of the spatial 
force resonances are obviously significant in gauge fixing. 
Such "gauge fixing" in the Eq. (\ref{AbrahamForce2}) 
to the London gauge \cite{mende}, postulated in the force expression (\ref{2AbrahamForceStationaryApproximation:a}) by 
London equations, can be clarified
by formal backgrounds of the electrodynamic nonlocality described in the Sec. \ref{Formal backgrounds}.  
Backgrounds of the spatial nonlocality  are closely relevant to
the
principles of gauge theories 
such as the configuration space, redundant degrees of freedom, classes. 
However, the Sec.\ref{Formal backgrounds} is addressed 
mainly to analyse the nonself-adjoint problems 
and further detailed formal comparisons of approaches presented
here,
and in gauge theories are beyond the scope of this study. 

The last notice 
also 
concerns the following conclusion. The equalities (\ref{AbrahamForce2}), (\ref{AbrahamForceStationaryApproximation:b}) and (\ref{2AbrahamForceStationaryApproximation:a}) demonstrate that 
finding the proper, empirically verified solution of the electrodynamic problem 
may require an introduction of the effective constitutive parameters (cf. Sec. \ref{Effective constitutive parameters}) or postulation of the additional phenomenological equations (cf. Sec. \ref{London equations and Abraham force}) 
depending on the kind of representation of the 
original problem operator. The reasons
for
the dependence on operator representation can be found also in the Sec. \ref{Formal backgrounds}. 
It 
has been argued that 
the combined vector manifold 
of the real three-dimensional linear space
and of the empiric physical fields 
transforms into the real 
nonlinear three-dimensional 
vector  manifold. 
This can affect the linear properties and commutativity of operators acting in this nonlinear manifold.
In turn,
the discussed dependence on operator representation can be the result of lost linearity and commutativity especially in the case of evaluating the indeterminate form in the zero frequency limit. It will be considered in 
details elsewhere.



\section{Conclusions}
\label{Conclusions}

The study
derived
the general frequency dependence of
the frequency-dependent surface impedance for the solutions corresponding to the spatially dispersive eigenvalues of the permittivity operator for conductors for all solutions including that for superconductors. It is shown that an incorporation of the spatial dispersion leads to an appearance  of the Meissner effect in perfect conductors in the same manner as in superconductors. Formal backgrounds of the electrodynamic nonlocality were deduced.
The 
surface impedance approximation has shown
comprehensibly
that the perfect conductivity  results
in
the Meissner effect, when
the correlation of
ponderomotive Abraham forces is taken into account as
an effect of the spatial dispersion.
It allows to
refuse the
introduction of the Meissner effect as the second necessary
attribute of the superconducting state in models
of the superconductivity 
when the first
necessary requirement
is the perfect conductivity.
This expanded conception is
promising for applications in novel
nanoelectronic
devices exploiting the
coherence,
nonlocality of the superconducting-like state and for search of
approaches to the problem of the room temperature superconductivity. The obtained results demonstrate that
finding the proper, empirically verified solution of the electrodynamic problem may require an introduction of the effective constitutive parameters or postulating the additional phenomenological equations in accordance with the kind of formal representation of the
original problem.

This research was supported by Grant of RFBR (18-02-00874).

\end{document}